\documentclass[tighten,twocolumn,appendixfloats,apj]{likeapj}
\pdfoutput=1
\usepackage{etoolbox}
\makeatletter

\patchcmd{\NAT@citex}
  {\@citea\NAT@hyper@{%
     \NAT@nmfmt{\NAT@nm}%
     \hyper@natlinkbreak{\NAT@aysep\NAT@spacechar}{\@citeb\@extra@b@citeb}%
     \NAT@date}}
  {\@citea\NAT@nmfmt{\NAT@nm}%
   \NAT@aysep\NAT@spacechar\NAT@hyper@{\NAT@date}}{}{}

\patchcmd{\NAT@citex}
  {\@citea\NAT@hyper@{%
     \NAT@nmfmt{\NAT@nm}%
     \hyper@natlinkbreak{\NAT@spacechar\NAT@@open\if*#1*\else#1\NAT@spacechar\fi}%
       {\@citeb\@extra@b@citeb}%
     \NAT@date}}
  {\@citea\NAT@nmfmt{\NAT@nm}%
   \NAT@spacechar\NAT@@open\if*#1*\else#1\NAT@spacechar\fi\NAT@hyper@{\NAT@date}}
  {}{}

  \makeatother
\usepackage{graphicx,bm,mathptmx,amsmath,float}

\hypersetup{linkcolor=blue,citecolor=blue,filecolor=cyan,urlcolor=blue,bookmarksnumbered=true,bookmarksopen=true}
\usepackage[all]{hypcap}
\usepackage{natbib}
\received{2018}
\revised{2018}
\accepted{2018}
\published{2018}

\shortauthors{Lazzarini et al.}

\newcommand{\ChandraPHAT}{Williams et al. 2018, submitted}

\newcommand{\tNuSTAR}{Wik et al. (2018, in prep.)}
\newcommand{\Zezas}{Zezas et al. (2018, in prep.)}
\newcommand{\Chandra}{\textit{Chandra}}
\newcommand{\Hubble}{\textit{HST}}
\newcommand{\NuSTARmission}{\textit{NuSTAR}}

\begin{document}
\title{Young Accreting Compact Objects in M31: The Combined Power of \textit{NuSTAR}, \textit{Chandra}, and \textit{Hubble}}
\author{M. Lazzarini}
\affiliation{Department of Astronomy, Box 351580, University of Washington, Seattle, WA 98195, USA}
\affiliation{Laboratory for X-ray Astrophysics, NASA Goddard Space Flight Center, Code 662, Greenbelt, MD 20771, USA}
\author{A. E. Hornschemeier}
\affiliation{Laboratory for X-ray Astrophysics, NASA Goddard Space Flight Center, Code 662, Greenbelt, MD 20771, USA}
\affiliation{Department of Physics and Astronomy, Johns Hopkins University, 3400 N. Charles Street, Baltimore, MD 21218, USA}
\author{B. F. Williams}
\affiliation{Department of Astronomy, Box 351580, University of Washington, Seattle, WA 98195, USA; mlazz@uw.edu}
\author{D. Wik}
\affiliation{Department of Physics and Astronomy, University of Utah, 201 James Fletcher Bldg., Salt Lake City, UT 84112, USA}
\affiliation{Laboratory for X-ray Astrophysics, NASA Goddard Space Flight Center, Code 662, Greenbelt, MD 20771, USA}
\author{N. Vulic}
\affiliation{Laboratory for X-ray Astrophysics, NASA Goddard Space Flight Center, Code 662, Greenbelt, MD 20771, USA}
\affiliation{Department of Astronomy and Center for Space Science and Technology (CRESST), University of Maryland, College Park, MD 20742, USA}
\author{M. Yukita}
\affiliation{Laboratory for X-ray Astrophysics, NASA Goddard Space Flight Center, Code 662, Greenbelt, MD 20771, USA}
\affiliation{Department of Physics and Astronomy, Johns Hopkins University, 3400 N. Charles Street, Baltimore, MD 21218, USA}
\author{A. Zezas}
\affiliation{Harvard-Smithsonian Center for Astrophysics, Cambridge, MA 02138, USA}
\author{A. R. Lewis}
\affiliation{Department of Astronomy, The Ohio State University, 140 West 18th Avenue, Columbus, OH 43210, USA}
\affiliation{Center for Cosmology and AstroParticle Physics, The Ohio State University, Columbus, OH 43210, USA}
\affiliation{Department of Astronomy, Box 351580, University of Washington, Seattle, WA 98195, USA; mlazz@uw.edu}
\author{M. Durbin}
\affiliation{Department of Astronomy, Box 351580, University of Washington, Seattle, WA 98195, USA; mlazz@uw.edu}
\author{A. Ptak}
\affiliation{Laboratory for X-ray Astrophysics, NASA Goddard Space Flight Center, Code 662, Greenbelt, MD 20771, USA}
\affiliation{Department of Physics and Astronomy, Johns Hopkins University, 3400 N. Charles Street, Baltimore, MD 21218, USA}
\author{A. Bodaghee}
\affiliation{Department of Chemistry, Physics and Astronomy, Georgia College and State University, Milledgeville, GA 31061, USA}
\author{B. D. Lehmer}
\affiliation{Department of Physics, University of Arkansas, 226 Physics Building, 825 West Dickson Street, Fayetteville, AR 72701, USA}
\author{V. Antoniou}
\affiliation{Harvard-Smithsonian Center for Astrophysics, Cambridge, MA 02138, USA}
\author{T. Maccarone}
\affiliation{Department of Physics and Astronomy, Box 41051, Science Building, Texas Tech University, Lubbock, TX 79409, USA}

\begin{abstract}
We present 15 high mass X-ray binary (HMXB) candidates in the disk of
M31 for which we are able to infer compact object type, spectral type
of the donor star, and age using multiwavelength observations
from \NuSTARmission, \Chandra, and the \textit{Hubble Space Telescope
(HST)}. The hard X-ray colors and luminosities from \NuSTARmission\
permit the tentative classification of accreting X-ray binary systems
by compact object type, distinguishing black hole from neutron star
systems. We find hard state black holes, pulsars, and non-magnetized
neutron stars associated with optical point source counterparts with
similar frequency. We also find nine non-magnetized neutron stars
coincident with globular clusters and an equal number of pulsars with
and without point source optical counterparts. We perform spectral
energy distribution (SED) fitting for the most likely optical
counterparts to the HMXB candidates, finding 7 likely high mass stars
and 1 possible red Helium burning star.  The remaining 7 HMXB optical
counterparts have poor SED fits, so their companion stars remain
unclassified. Using published star formation histories, we find that
the majority of HMXB candidates --- X-ray sources with UV-bright point
source optical counterpart candidates --- are found in regions with
star formation bursts less than 50 Myr ago, with 3 associated with
young stellar ages ($<$10 Myr). This is consistent with similar
studies of HMXB populations in the Magellanic Clouds, M33, NGC 300,
and NGC 2403.
\end{abstract}

\keywords{galaxies: individual (M31), X-rays: binaries, NuSTAR, \Chandra, Hubble Space Telescope}

\section{Introduction}\label{sec1}
The production of extragalactic X-ray binaries (XRBs) is closely related to properties of the galaxies in which they form, such as the star formation rate \citep[e.g.][]{Ranalli2003,Gilfanov2004,Mineo2012}, stellar mass \citep{Lehmer2010}, and metallicity \citep{Basuzych2013,Brorby2016}. Population studies of XRBs probe the production of these compact objects and their relationship to their host galaxy properties. However, the fundamental properties of XRBs, such as the compact object type and the physical properties of the donors, have remained difficult to determine given the limited information contained in the 0.5-10 keV energy range covered by soft X-ray telescopes such as \Chandra\ and \textit{XMM-Newton}. Broadening the observed energy range to include data from the near IR (\textit{Hubble Space Telescope (HST)}) through hard X-rays (\NuSTARmission) allows us to determine the compact object type, the physical properties of the donors, and place constraints on the age of XRBs using star formation histories for their surrounding stellar populations.
\par Conducting a galaxy-wide study of high mass X-ray binaries (HMXBs) in connection to their star forming environments is challenging in the Milky Way due to distance uncertainties, but there has been some successful work in this area \citep{Grimm2002}. Additionally, \cite{Bodaghee2012} used the spatial correlation between HMXBs and OB associations in the Milky Way to determine ages of the systems. It is expected that 5-10 Myr elapses between the formation of a high mass star and the supernova, which forms the compact object in HMXBs \citep{Schaller1992,Linden2010}. Thus, an HMXB cannot migrate far from its birthplace, allowing its spatial correlation with an OB association to be used to constrain its age. \citet{Bodaghee2012} determined the time from supernova through the HMXB phase (the ``kinematic age'') using the spatial correlation between HMXB candidates and OB associations.  They found that most systems have kinematic ages of $\sim 4$ Myr.
\par Detailed studies of XRBs in extragalactic star forming environments have been done previously in the Small Magellanic Cloud (SMC), Large Magellanic Cloud (LMC), NGC 300, NGC 2403, and M33. In the SMC, B\emph{e}/X-ray binaries are found in regions with star forming bursts 25-60 Myr ago \citep{SMC}. In the LMC, HMXBs are found in areas with considerably more recent star formation, between 6 and 25 Myr ago \citep{LMC}. In NGC 300 and NGC 2403, HMXB candidates have been found in regions with surrounding stellar populations between 20 and 70 Myr old \citep{Williams2013} with a peak at 40-55 Myr, agreeing with ages in the SMC. In M33 a similar set of peaks is seen in the HMXB age distribution (Garofali et al. 2018, submitted). These ages suggest two potential formation channels: one that operates on the timescale of B-star evolution ($\sim$50 Myr) and the other that operates much more promptly.


\par To better connect the properties of the XRBs themselves to their parent populations, classifying the compact object in the system is critical. However, compact object characterization can be difficult because there are currently very limited methods available. If a low-mass XRB has an observed Type-I X-ray burst, its compact object may be classified as a neutron star \citep[e.g.,][]{Type1}. Black holes can be classified as such if the mass of the compact object can be confirmed using the orbital period and mass of the companion \citep{Orosz}, but stellar companion orbits are not always available, especially for extragalactic XRBs.
\par The hard X-ray coverage of \NuSTARmission\ allows compact objects to be tentatively classified using their X-ray properties. With X-ray observations that cover the hard band (4-25 keV), compact objects can be classified as neutron stars or black holes based on a combination of their X-ray colors and luminosities \citep[Zezas et al. 2018, in prep.;][]{Zezas2014,Wik2014,Yukita2016}. This can be done because neutron stars always have hard emission associated with matter accreting onto the surface, while black holes do not and are dominated by the disk emission properties \citep{Maccarone2016}. Thus, an XRB's hard X-ray colors distinguish between neutron star and black hole systems. Techniques involving hard X-rays are more indirect but are critical for expanding our tool kit for classifying X-ray sources as black holes or neutron stars.
\par Andromeda (M31), the nearest spiral galaxy to the Milky Way, is one of the best systems for studying X-ray binary populations in the context of their star forming environments because of its proximity and the large number of multiwavelength data sets available. Observations with the sensitivity to detect faint point sources extend from near infrared wavelengths up to hard X-rays (E $\lesssim$ 50 keV) \citep[e.g.,][]{WilliamsPHAT,Yukita,Maccarone2016,Vulic2014,Vulic}.
\par There has recently been a major improvement in the X-ray coverage of Andromeda owing to two powerful and deep surveys by \NuSTARmission\ and \Chandra, both taken in 2015. \NuSTARmission\ observed an $\sim 750$ arcmin$^{2}$ area of M31 at a depth of $\sim$1.4 Ms \citep[2018, in prep.]{WikHEAD,WikAAS}. A \Chandra\ Large Project survey covered $\sim 1800$ arcmin$^{2}$ to a depth of 50 ks (ChandraPHAT; \ChandraPHAT). 
\par We pair these X-ray observations with existing near-infrared to ultraviolet observations from Hubble \citep[PHAT;][]{Dalcanton,WilliamsPHAT} to study hard X-ray emitting compact objects and their optical counterparts in the context of their star forming environments. A total area of $\sim 570$ arcmin$^{2}$ is covered by all three telescopes. This area comprises approximately 6\% of the $D_{25}$ area of M31.
\par The maturity of the PHAT project means that invaluable secondary data products are available to characterize the star forming environments around X-ray sources. For example, \cite{Lewis} spatially mapped the recent star formation history of M31 and \cite{Gregersen} mapped the metallicity distribution. Both properties allow the X-ray binary population to be put in the context of its environment. The Bayesian Extinction and Stellar Tool (BEAST) by \cite{BEAST} can fit the spectral energy distribution (SED) of individual stars in the disk of M31. The BEAST code provides a powerful tool for understanding the physical characteristics of the companion star in an XRB. Additionally, M31 allows us to study XRB populations in their environments without the uncertainties in the distance to each system that plague such studies in the Milky Way.
\par In this paper we use the multiwavelength coverage from \NuSTARmission, \Chandra, and \Hubble\ to investigate the HMXB population in the northern disk of M31. In Section \ref{sec2}, we describe the three data sets used in this study: \NuSTARmission\ observations, two sets of \Chandra\ observations, and reduced \Hubble\ photometry and imaging from the PHAT survey. We describe the methods used to match sources between the three data sets in Section \ref{sec3}. In Section \ref{sec4}, we describe our results: how \NuSTARmission\ sources were classified using their X-ray colors and luminosities, how we determined ages for HMXB candidates using spatially resolved star formation histories, and the SED fitting used to determine spectral types for companion stars in HMXB candidates. In Section \ref{sec5}, we discuss our results in the context of previous studies and in Section \ref{sec6}, we provide a brief summary of our results.
\par We assume a  Galactic column density, $N_{H}=7 \times 10^{20}\ cm^{-2}$, and a photon index, $\Gamma = 1.7$ \citep{Stiele}, to convert count rates to absorbed energy flux. We assume a distance of 776 kpc to M31 \citep{Dalcanton} for luminosity calculations.
\section{Data}\label{sec2}
\begin{figure*}
\centering
\includegraphics[width=0.95\textwidth]{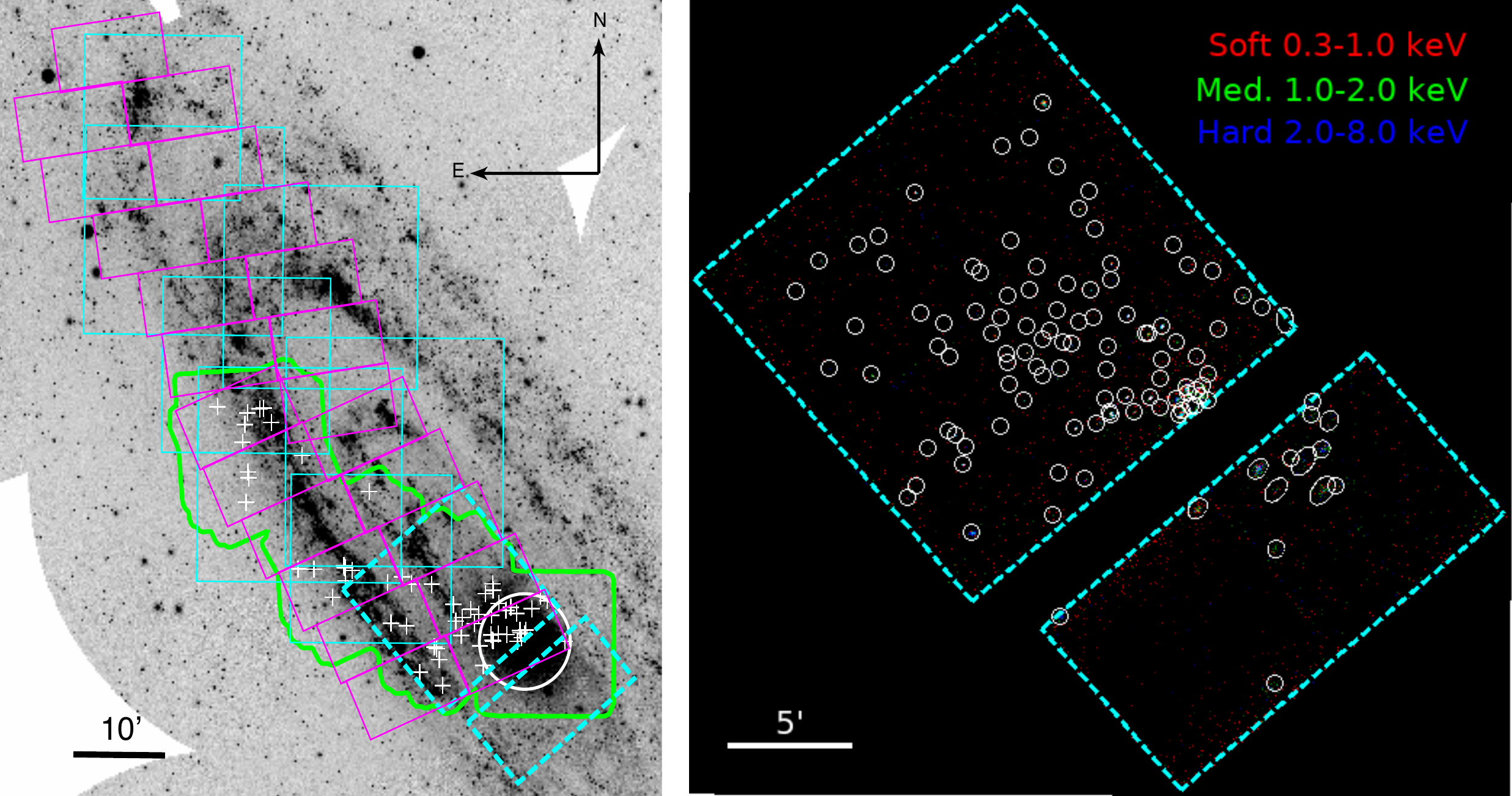}
\caption{Near UV image of M31 from the Galaxy Evolution Explorer (GALEX) (left) \citep{GALEX} and three color X-ray image of \Chandra\ Field A (right), see Section \ref{chandra_data} for more information on Field A data reduction. Magenta regions outline the area observed by PHAT. Green outlines the \NuSTARmission\ observed region and cyan outlines the area observed by \Chandra, with solid lines indicating ChandraPHAT observations and dashed lines indicating Field A observations. In the UV image, the 64 sources observed by \NuSTARmission\ and \Chandra\ that fall within the PHAT footprint are marked with white crosses and a 5$^\prime$ circle outlines the densest inner bulge region of M31. In the X-ray image, all sources detected by wavdetect within Field A are marked with white ellipses (including sources that do not match to \NuSTARmission\ sources or are outside the PHAT footprint, and thus are not presented in our sample in Table \ref{chan_nustar}).}
\label{m31}
\end{figure*}

In this study we employ data from \NuSTARmission, \Chandra, and \Hubble. We now describe each data set in more detail below. For an overview of the area observed by each telescope, see Figure \ref{m31}.
\subsection{\NuSTARmission\ Data}
\NuSTARmission\ source catalogs and source classifications come from \citet[2018, in prep.]{WikHEAD,WikAAS}. Observations were taken between October and February 2015 covering the area outlined in green in Figure \ref{m31} with an average exposure time of 1.4 Ms and were reduced using the \verb!nupipeline! software. Sources previously observed with \Chandra\ were used for astrometric alignment. For detailed information on data reduction, source detection, and source classification please see \tNuSTAR, which presents the entire \NuSTARmission\ M31 survey. The \NuSTARmission\ observations cover the nucleus and inner disk regions of M31 at an energy range of 4-25 keV using the 4-6 keV, 6-12 keV, and 12-25 keV energy bands. The completeness of the \NuSTARmission\ observations starts to fall off at a luminosity of $\sim 3\times 10^{36}\ erg\ s^{-1}$ and reaches zero at $\sim 2\times 10^{36}\ erg\ s^{-1}$.

\renewcommand{\arraystretch}{0.95}
\setlength{\tabcolsep}{2.5pt}
\startlongtable
\begin{deluxetable*}{ccllcccclcc}
\tablecaption{All sources observed by \NuSTARmission\ and \Chandra\ within the PHAT footprint \label{chan_nustar}}
\tabletypesize{\scriptsize}
\tablehead{
\colhead{\NuSTARmission} &
\colhead{\Chandra} &
\colhead{\Chandra} &
\colhead{\Chandra} &
\colhead{\Chandra} &
\colhead{\Chandra} &
\colhead{Theta [$^\prime$]} &
\colhead{\Chandra\ Flux} &
\colhead{Stiele} &
\colhead{Stiele} &
\colhead{PHAT}
\\
\colhead{ID} &
\colhead{Catalog Name} &
\colhead{RA} &
\colhead{Dec} &
\colhead{RA err [$^{\prime \prime}$]} &
\colhead{Dec err [$^{\prime \prime}$]} &
\colhead{} &
\colhead{(0.35-8.0 keV)} &
\colhead{ID} &
\colhead{Class.} &
\colhead{Cpt.}
\\
\colhead{} &
\colhead{} &
\colhead{} &
\colhead{} &
\colhead{} &
\colhead{} &
\colhead{} &
\colhead{[$\times 10^{-13}\ \rm erg\ cm^{-2}\ s^{-1}$]} &
\colhead{} &
\colhead{} &
\colhead{} 
}
\startdata
19 & $004220.96+411520.3$ & 10.587329 & 41.255773 & 2.0 & 2.0 & 13.0 & $4.2^{+0.4}_{-0.4}$ & \nodata & \nodata & n \\
24 & 004231.27+411937.5 & 10.630292 & 41.327141 & 2.0 & 2.0 & 10.0 & $1.9^{+0.3}_{-0.2}$ & 923 & GlC & c \\
26 & 004235.20+412005.0 & 10.646687 & 41.334782 & 1.0 & 0.9 & 9.5 & $3.4^{+0.3}_{-0.3}$ & 952 & $\langle$hard$\rangle$ & n \\
27 & 004240.31+411845.3 & 10.667979 & 41.312682 & 2.0 & 1.0 & 8.6 & $0.90^{+0.15}_{-0.14}$ & 972 & $\langle$hard$\rangle$ & p \\
41 & 004243.81+411631.0 & 10.682538 & 41.275371 & 0.7 & 0.4 & 8.3 & $13.0^{+0.7}_{-0.6}$ & 1005 & $\langle$XRB$\rangle$ & n \\
43 & 004244.27+411607.6 & 10.684498 & 41.268868 & 0.7 & 0.4 & 8.4 & $19.0^{+0.9}_{-0.9}$ & 1010 & $\langle$XRB$\rangle$ & n \\
44 & 004246.19+411543.2 & 10.692552 & 41.262086 & 2.0 & 2.0 & 8.2 & $0.76^{+0.25}_{-0.21}$ & \nodata & \nodata & n \\
45 & 004246.97+411615.3 & 10.695737 & 41.271019 & 0.6 & 0.4 & 7.9 & $13.0^{+0.6}_{-0.6}$ & 1023 & $\langle$AGN$\rangle$ & n \\
46 & 004247.18+411628.0 & 10.696608 & 41.274542 & 0.6 & 0.3 & 7.8 & $26.0^{+0.8}_{-0.8}$ & 1024 & $\langle$XRB$\rangle$ & n \\
47 & 004248.56+411520.8 & 10.702332 & 41.255889 & 0.6 & 0.3 & 8.0 & $51.0^{+1.0}_{-1.0}$ & 1036 & $\langle$XRB$\rangle^{a}$ & n \\
54 & 004249.22+411815.5 & 10.70509 & 41.304395 & 0.8 & 0.5 & 6.9 & $1.5^{+0.2}_{-0.1}$ & 1041 & $\langle$hard$\rangle$ & p \\
55 & 004252.53+411854.0 & 10.718888 & 41.315082 & 0.5 & 0.2 & 6.3 & $19.0^{+0.5}_{-0.5}$ & 1060 & $\langle$XRB$\rangle$ & n \\
65 & 004254.93+411602.8 & 10.728903 & 41.267544 & 0.5 & 0.2 & 6.6 & $16.0^{+0.5}_{-0.5}$ & 1075 & $\langle$XRB$\rangle^{a}$ & n \\
57 & 004255.19+411835.4 & 10.729964 & 41.309922 & 0.7 & 0.5 & 5.8 & $0.83^{+0.12}_{-0.10}$ & 1078 & $\langle$hard$\rangle$ & n \\
57 & 004255.60+411834.5 & 10.731708 & 41.309676 & 0.7 & 0.4 & 5.7 & $0.91^{+0.12}_{-0.11}$ & 1078 & $\langle$hard$\rangle$ & c \\
59 & 004259.66+411918.9 & 10.748579 & 41.322021 & 0.4 & 0.2 & 4.9 & $11.0^{+0.4}_{-0.4}$ & 1102 & GlC & c \\
68 & 004259.88+411605.3 & 10.749525 & 41.268225 & 0.5 & 0.2 & 5.8 & $10.0^{+0.4}_{-0.4}$ & 1103 & GlC & c \\
70 & 004302.94+411522.2 & 10.762284 & 41.256263 & 0.5 & 0.2 & 5.8 & $6.4^{+0.3}_{-0.3}$ & 1116 & GlC & c \\
77 & 004303.03+412041.6 & 10.762621 & 41.344983 & 0.6 & 0.4 & 4.5 & $0.50^{+0.09}_{-0.07}$ & 1115 & $\langle$hard$\rangle$ & n \\
70 & 004303.23+411527.3 & 10.763472 & 41.257677 & 0.5 & 0.2 & 5.7 & $13.0^{+0.4}_{-0.4}$ & 1116 & GlC & n \\
78 & 004303.29+412121.5 & 10.763713 & 41.35607 & 0.5 & 0.2 & 4.7 & $2.3^{+0.2}_{-0.2}$ & 1118 & GlC & c \\
60 & 004303.87+411804.5 & 10.766124 & 41.301336 & 0.4 & 0.2 & 4.3 & $6.0^{+0.3}_{-0.3}$ & 1122 & GlC & c \\
71 & 004304.25+411600.7 & 10.767706 & 41.266967 & 0.7 & 0.5 & 5.2 & $0.59^{+0.10}_{-0.09}$ & 1124 & $\langle$GlC$\rangle$ & n \\
79 & 004307.51+412019.4 & 10.781315 & 41.338806 & 0.5 & 0.3 & 3.6 & $0.58^{+0.09}_{-0.08}$ & 1137 & $\langle$GlC$\rangle$ & c \\
80 & 004308.62+411248.0 & 10.785948 & 41.213448 & 0.8 & 0.5 & 7.2 & $3.2^{+0.3}_{-0.3}$ & 1146 & XRB & p \\
81 & 004310.62+411451.0 & 10.794248 & 41.247599 & 0.4 & 0.2 & 5.2 & $23.0^{+0.6}_{-0.6}$ & 1157 & GlC & c \\
82 & 004311.37+411809.3 & 10.797389 & 41.302675 & 0.5 & 0.2 & 2.9 & $0.78^{+0.10}_{-0.09}$ & 1160 & $\langle$hard$\rangle$ & n \\
85 & 004313.88+411711.5 & 10.807835 & 41.286615 & 0.7 & 0.5 & 3.0 & $0.13^{+0.05}_{-0.04}$ & \nodata & \nodata & n \\
86 & 004316.10+411841.2 & 10.817115 & 41.311543 & 0.4 & 0.2 & 1.9 & $0.35^{+0.06}_{-0.07}$ & 1180 & $\langle$XRB$\rangle$ & p \\
88 & 004321.07+411750.2 & 10.837815 & 41.297389 & 0.4 & 0.1 & 1.7 & $1.1^{+0.1}_{-0.1}$ & 1203 & $\langle$hard$\rangle$ & p \\
87 & 004321.48+411556.5 & 10.839501 & 41.265805 & 0.7 & 0.4 & 3.4 & $0.21^{+0.05}_{-0.06}$ & \nodata & \nodata & p \\
89 & 004324.84+411726.9 & 10.853509 & 41.290917 & 0.4 & 0.2 & 1.8 & $0.47^{+0.08}_{-0.07}$ & 1216 & $\langle$hard$\rangle$ & n \\
90 & 004326.33+411911.4 & 10.859718 & 41.31994 & 0.4 & 0.1 & 0.11 & $0.38^{+0.09}_{-0.07}$ & 1224 & $\langle$AGN$\rangle$ & g \\
91 & 004332.38+411040.9 & 10.884951 & 41.178136 & 0.7 & 0.4 & 8.7 & $15.0^{+0.5}_{-0.5}$ & 1253 & $\langle$hard$\rangle^{a}$ & n \\
92 & 004334.33+411323.1 & 10.893064 & 41.223187 & 0.5 & 0.3 & 6.1 & $5.4^{+0.3}_{-0.3}$ & 1261 & $\langle$hard$\rangle$ & n \\
93 & 004335.91+411433.0 & 10.899635 & 41.2426 & 0.8 & 0.6 & 5.1 & $0.41^{+0.09}_{-0.07}$ & 1262 &  & p \\
94 & 004337.28+411443.1 & 10.905322 & 41.245424 & 0.4 & 0.2 & 5.0 & $16.0^{+0.4}_{-0.4}$ & 1267 & GlC & c \\
95 & 004339.06+412116.7 & 10.912737 & 41.354885 & 0.7 & 0.7 & 7.2 & $0.73^{+0.07}_{-0.06}$ & \nodata & \nodata & p \\
96 & 004345.83+411203.7 & 10.940976 & 41.201128 & 2.0 & 2.0 & 8.1 & $0.39^{+0.11}_{-0.09}$ & 1298 & $\langle$hard$\rangle$ & n \\
99 & 004350.76+412117.4 & 10.961516 & 41.355033 & 0.4 & 0.4 & 5.1 & $0.65^{+0.06}_{-0.05}$ & 1319 & $\langle$hard$\rangle$ & p \\
97 & 004353.65+411654.6 & 10.973526 & 41.282044 & 0.4 & 0.4 & 7.6 & $6.6^{+0.2}_{-0.2}$ & 1327 & $\langle$GlC$\rangle$ & n \\
98 & 004356.43+412202.3 & 10.985126 & 41.367503 & 0.4 & 0.4 & 3.8 & $0.45^{+0.05}_{-0.05}$ & 1340 & GlC & c \\
105 & 004402.72+411711.3 & 11.011322 & 41.28666 & 1.0 & 1.0 & 6.5 & $0.18^{+0.04}_{-0.03}$ & \nodata & \nodata & n \\
101 & 004404.75+412126.5 & 11.019799 & 41.35756 & 0.3 & 0.3 & 2.7 & $0.75^{+0.06}_{-0.06}$ & 1373 & $\langle$AGN$\rangle$ & p \\
102 & 004416.02+413057.3 & 11.066667 & 41.516147 & 0.4 & 0.4 & 4.3 & $1.7^{+0.1}_{-0.1}$ & 1420 & XRB & n \\
103 & 004425.73+412241.8 & 11.107221 & 41.378442 & 0.3 & 0.3 & 2.0 & $0.22^{+0.03}_{-0.03}$ & 1450 & $\langle$hard$\rangle$ & p \\
104 & 004429.57+412135.1 & 11.123203 & 41.359913 & 0.3 & 0.3 & 3.1 & $18.0^{+0.3}_{-0.3}$ & 1463 & GlC & c \\
105 & 004429.73+412257.4 & 11.123878 & 41.382771 & 0.5 & 0.5 & 2.6 & $0.09^{+0.02}_{-0.02}$ & \nodata & \nodata & n \\
109 & 004430.16+412301.1 & 11.125694 & 41.383802 & 0.6 & 0.6 & 2.7 & $0.04^{+0.02}_{-0.01}$ & \nodata & \nodata & g \\
105 & 004430.45+412310.1 & 11.126901 & 41.3863 & 0.4 & 0.4 & 2.8 & $0.21^{+0.03}_{-0.03}$ & 1468 & $\langle$hard$\rangle$ & n \\
106 & 004437.08+411951.1 & 11.154504 & 41.331024 & 0.5 & 0.5 & 5.3 & $0.42^{+0.05}_{-0.05}$ & 1488 & $\langle$hard$\rangle$ & g \\
100 & 004448.13+412247.4 & 11.200545 & 41.379973 & 0.7 & 0.7 & 6.1 & $0.27^{+0.04}_{-0.04}$ & 1525 & $\langle$hard$\rangle$ & p \\
110 & 004455.53+413440.3 & 11.231167 & 41.57808 & 0.3 & 0.3 & 2.7 & $0.45^{+0.05}_{-0.04}$ & 1547 & $\langle$AGN$\rangle$ & n \\
108 & 004457.39+412247.9 & 11.239115 & 41.380094 & 0.7 & 0.7 & 7.8 & $1.0^{+0.1}_{-0.1}$ & 1553 & $\langle$XRB$\rangle$ & n \\
111 & 004513.82+413806.4 & 11.307524 & 41.635323 & 0.8 & 0.8 & 6.1 & $0.24^{+0.04}_{-0.04}$ & 1598 & $\langle$hard$\rangle$ & g \\
112 & 004518.39+413936.0 & 11.326586 & 41.66018 & 0.5 & 0.5 & 4.4 & $0.28^{+0.04}_{-0.04}$ & 1611 & $\langle$hard$\rangle$ & p \\
113 & 004520.74+413932.1 & 11.336316 & 41.659109 & 0.6 & 0.6 & 4.3 & $0.14^{+0.03}_{-0.03}$ & \nodata & \nodata & g \\
117 & 004526.86+413216.8 & 11.361729 & 41.538161 & 0.6 & 0.6 & 5.2 & $0.27^{+0.04}_{-0.03}$ & 1631 & $\langle$AGN$\rangle$ & n \\
118 & 004527.34+413253.5 & 11.363743 & 41.548363 & 0.4 & 0.4 & 5.4 & $2.2^{+0.1}_{-0.1}$ & 1634 & $\langle$hard$\rangle$ & g \\
114 & 004527.89+413904.9 & 11.366179 & 41.651539 & 0.4 & 0.4 & 4.3 & $0.43^{+0.05}_{-0.04}$ & 1635 & $\langle$hard$\rangle$ & p \\
119 & 004528.29+412943.4 & 11.367681 & 41.495538 & 0.4 & 0.4 & 6.0 & $1.9^{+0.1}_{-0.1}$ & 1636 & $\langle$hard$\rangle$ & p \\
115 & 004529.35+413751.6 & 11.37223 & 41.631176 & 1.0 & 1.0 & 5.6 & $0.06^{+0.02}_{-0.02}$ & \nodata & \nodata & g \\
116 & 004530.65+413559.8 & 11.377557 & 41.600135 & 0.9 & 0.9 & 7.1 & $0.38^{+0.06}_{-0.05}$ & 1643 & $\langle$hard$\rangle$ & g \\
120 & 004545.57+413941.5 & 11.439867 & 41.661701 & 0.3 & 0.3 & 4.7 & $71.0^{+0.6}_{-0.6}$ & 1692 & GlC & c\\ 
\enddata
\tablecomments{List of all \NuSTARmission-\Chandra\ sources within the PHAT footprint. Optical counterparts to X-ray sources are listed in the \textit{PHAT Cpt.} column: \textit{g} = galaxy, \textit{n} = no optical counterpart, \textit{c} = cluster and \textit{p} = point source. See Section \ref{opt_counterparts} for further discussion of optical counterpart determination. Stiele ID and classifications from \cite{Stiele}. \newline$^{a}$Sources have updated Stiele classifications from \citet{Stiele2018}.}
\end{deluxetable*}

\renewcommand{\arraystretch}{1.}

\subsection{\Chandra\ Data}\label{chandra_data}
\Chandra\ data used in this study are comprised of two data sets: the ChandraPHAT data, a \Chandra\ Large Project survey by B. Williams et al. (2018, submitted), and one additional \Chandra\ field (obsid 18046, P.I. Hornschemeier), hereby referenced as Field A, that was reduced for this paper. The ChandraPHAT data set is comprised of 7 \Chandra\ pointings, each with a depth of about 50 ks. The Field A data is made of one \Chandra\ pointing with a depth of 25 ks. The ChandraPHAT observations were taken in October 2015 and the Field A observation was taken in August 2016. The completeness in the ChandraPHAT field starts to drop at a luminosity of $\sim 3\times 10^{35}\ erg\ s^{-1}$ and reaches zero at $\sim 5\times 10^{34}\ erg\ s^{-1}$. The completeness in Field A starts to drop at $\sim 7\times 10^{35}\ erg\ s^{-1}$ and reaches zero at $\sim 1\times 10^{35}\ erg\ s^{-1}$.
\par Field A is centered at (RA,Dec)=(00:43:29.30, +41:18:21.20) and was designed to overlap with \NuSTARmission\ Field A in the observations by \tNuSTAR. For a detailed description of the data reduction for \Chandra\ sources within the ChandraPHAT footprint, see Williams et al. (2018, submitted). Sections \ref{detection} and \ref{astrometry} contain a detailed description of the reduction of the \Chandra\ Field A data, which follows the methodology in Williams et al. (2018, submitted).
\subsubsection{Source Detection}\label{detection}
We generated the initial source list using the \verb!wavdetect! tool of CIAO version 4.9 and CALDB version 4.7.7 \citep{CIAO}. We produced the source image and exposure map using the CIAO command \verb!fluximage! and created the PSF map using the CIAO tool \verb!mkpsfmap! with the standard parameters, energy=1.4967 and ecf=0.393. We then ran \verb!wavdetect! using the source image and PSF map to create a source list. The wavelet scales were set to 1.0, 2.0, 3.0, 8.0, and 16.0 pixels.
\par We ran ACIS extract (AE) version 2016sept22 \citep{Broos2010} on the source list output from \verb!wavdetect!. We followed Section 3.2 of the AE users guide to prepare the event files, aspect histogram file, aspect solution file, and mask file for AE source extraction. With the input source list from \verb!wavdetect!, we extracted sources with energy limits of 0.35-8.0 keV. See Table \ref{chan_nustar} for the positions, off-axis angle, and net counts output by AE.
\begin{figure}
    \centering
    \includegraphics[width=0.47\textwidth]{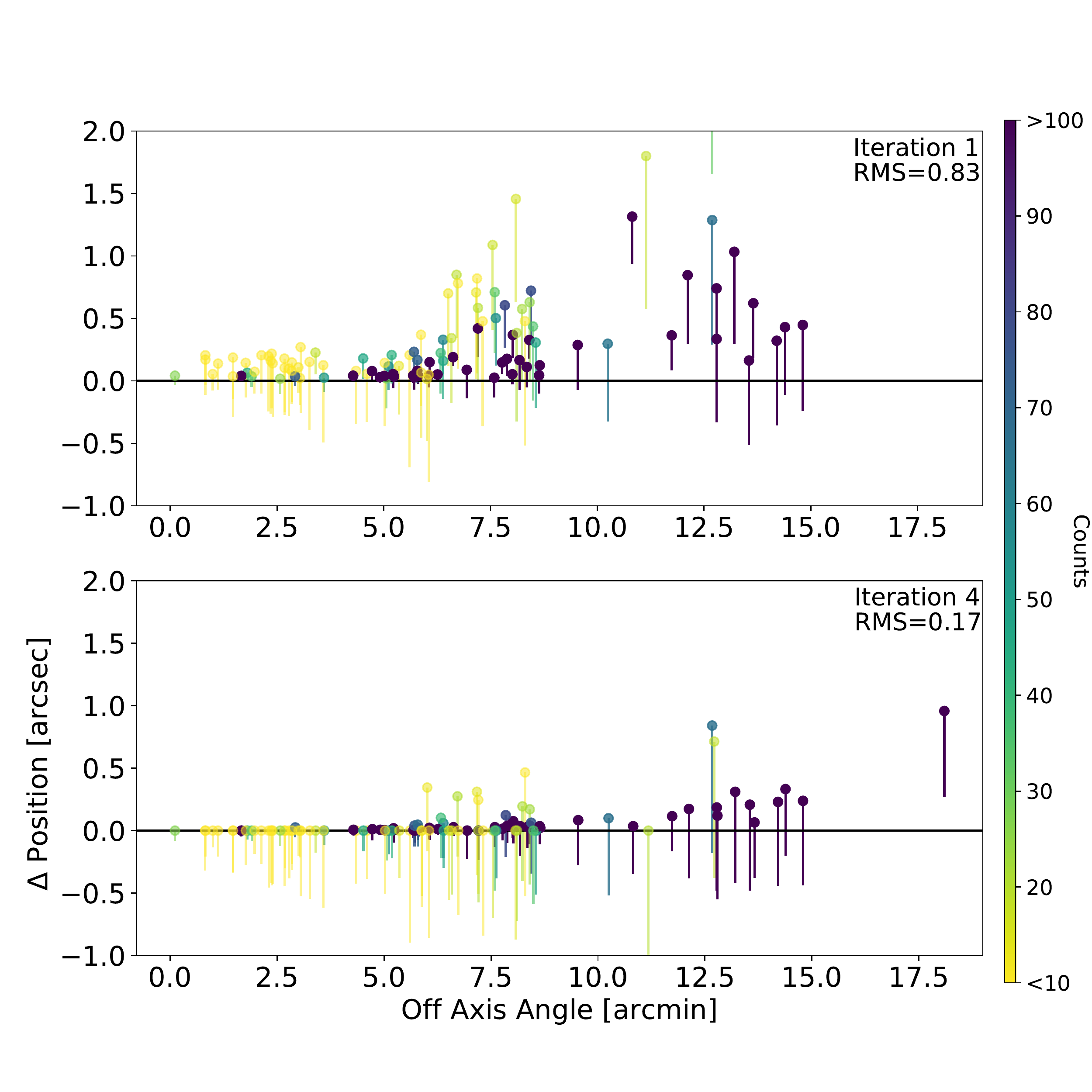}
    \caption{Results of ACIS extract iteration to improve \Chandra\ source positions in Field A observations (see Section \ref{detection}). The top panel shows the change in source position between the input source list from wavdetect and the output AE positions. The bottom panel shows the change between the input and output source positions for the fourth iteration of AE. Note that most of the sources with an off-axis angle $>10^\prime$ have more than 100 counts. These belong to the nucleus of M31, which is far off axis in the Field A observations and has a high source density. The source with an off axis angle of $\sim$ 17.5$^\prime$ in the bottom panel does not show in the top panel because it has a $\Delta$ position $>$ 2$^{\prime \prime}$ between the input position from wavdetect and the output position from AE.}
    \label{ae_iterate}
\end{figure}
\par We iterated AE four times in order to obtain the most precise positions. The first AE run used the output positions from CIAO \verb!wavdetect! as the initial positions. Each subsequent AE run used the data mean position output from the previous run as the initial positions, this was repeated until the input and output data mean positions converged (Figure \ref{ae_iterate}). 
\subsubsection{Astrometric Alignment}\label{astrometry}
Once we obtained precise positions for the \Chandra\ sources with ACIS extract, we aligned  the Field A observations to the PHAT data set \citep{Dalcanton}. \NuSTARmission\ data were previously aligned to the ChandraPHAT data by \tNuSTAR.
\par We used 8 bright globular clusters that were detected at optical wavelengths by \Hubble\ and X-ray wavelengths by \Chandra\ for astrometric alignment. Clusters were identified by visual inspection of the PHAT imaging. First, we measured the centroids of the clusters using the \verb!centroid_1dg! tool in the \verb!photutils! (v0.4) Python package. To find the astrometric solution, we used the CIAO tool \verb!wcs_match! that aligns the \Chandra\ sources in a given image to the measured cluster positions from the PHAT images and outputs an astrometric solution. The parameters used in \verb!wcs_match! were radius=5, residlim=0, residtype=0, and residfac=25 using the description in \citet{Vulic} as a guide. The CIAO tool \verb!wcs_update! was used to updated the header of the \Chandra\ images and update the RA and Dec of the measured source positions.
\par Positional errors were calculated using the net counts in the 0.35-8.0 keV band and the off axis angle using the formula in \citet{Hong2005}, listed as Equation 5. Instead of using the 0.25$^{\prime \prime}$ baseline error in that equation, we added the residuals from astrometric alignment to PHAT, which were 0.29$^{\prime \prime}$ in RA and 0.03$^{\prime \prime}$ in Dec.
\par We merged the Field A catalog prepared for this paper with the ChandraPHAT catalog from Williams et al. (2018, submitted) for source matching with the \NuSTARmission\ source catalog by Wik et al. (2018, in prep.) and PHAT, detailed in Section \ref{sec3}.
\subsection{\Hubble\ Data: Panchromatic \textit{Hubble} Andromeda Treasury}
The \Hubble\ data come from the published Panchromatic \textit{Hubble} Andromeda Treasury (PHAT) dataset \citep{Dalcanton}. The PHAT survey imaged roughly a third of the disk of M31 in six \Hubble\ filters ranging from near infrared to ultraviolet wavelengths: F160W, F110W, F814W, F475W, F336W, and F275W (central $\lambda$ = 1.150 $\mu$, 1.545 $\mu$, 8353 \AA, 4750 \AA, 3375 \AA, 2750 \AA). We use published photometry catalogs by \citet{WilliamsPHAT} for optical counterpart analysis. PHAT observations were taken in 2010 and 2011. The PHAT data have a limiting F475W magnitude of $\sim 28$ in the outer disk, and $\sim 25$ in the more crowded inner disk region.
\section{Source Matching Between Data Sets}\label{sec3}
We first matched \NuSTARmission\ sources to \Chandra\ to find precise positions. We then used the \Chandra\ positions to identify optical counterparts in the PHAT data.
\subsection{Source Matching Between \NuSTARmission\ and \Chandra}\label{nust_chan_matching}
We identified 60 \NuSTARmission\ sources with positions inside the PHAT observed area that positionally match to \Chandra\ sources. These 60 sources have 64 associated \Chandra-detected X-ray sources. We cross-matched \NuSTARmission\ and  \Chandra\ sources within 10$^{\prime \prime}$ so we could use the more precise \Chandra\ positions to identify optical counterparts. 
\par We chose a 10$^{\prime \prime}$ match radius to account for the 9$^{\prime \prime}$ full width at half-maximum of the \NuSTARmission\ point spread function (PSF) and the $\sim$0.5$^{\prime \prime}$ mean \Chandra\ positional errors for sources in our sample. We measured the \Chandra\ exposure time at the position of each detected \NuSTARmission\ source to confirm that if a source was observed by both telescopes, there was a match.
\par In order to quantify the confidence level of these matches, we investigated the false match probability between \NuSTARmission\ and \Chandra. To do this, we adjusted the \NuSTARmission\ source positions for the full 121-source \NuSTARmission\ M31 catalog (Wik et al. 2018, in prep.) by 10$^{\prime \prime}$ in both RA and Dec. We performed this adjustment four times, using all permutations of adding and subtracting 10$^{\prime \prime}$ from the RA and Dec of \NuSTARmission\ sources. We re-matched the \NuSTARmission\ and \Chandra\ source positions each time to see how many \Chandra\ sources matched to the adjusted \NuSTARmission\ source positions. We found an average of 5 matches between the adjusted \NuSTARmission\ source positions and the \Chandra\ source positions. Out of 121 \NuSTARmission\ sources, this equals a false match probability of 4.1\%. This means that 2-3 of the \NuSTARmission\ sources in our sample could have false matches to \Chandra\ sources.
\par We note that while our sample covers an area $\sim$ 75\% of the \NuSTARmission\ total observed area, it only contains $\sim$ 50\% of the \NuSTARmission\ sources in the full 121 source catalog. This is because our sample only contains sources observed with \NuSTARmission, \Chandra, and \Hubble, which excludes part of the bulge of M31, in an area with high \NuSTARmission\ source density.
\par There are five \NuSTARmission\ sources that are within the \Chandra\ Field A footprint that were not detected by \Chandra. Given that the \Chandra\ Field A and \NuSTARmission\ observations were taken a year apart, we believe this discrepancy is due to variability. 
\par There are three \NuSTARmission\ sources whose positions are compatible with multiple \Chandra\ sources. This is not surprising as \NuSTARmission\ can blend \Chandra\ sources together because of its large PSF. The PSF of \NuSTARmission\ has a core with a full width at half-maximum of 18$^{\prime \prime}$ and a half-power diameter of 58$^{\prime \prime}$ \citep{Harrison2013}. This is quite large compared to the \Chandra\ PSF, which is $\sim$0.5$^{\prime \prime}$ on-axis to $\sim$10$^{\prime \prime}$ at the edge of the field \citep{Williams2004}.
\par \NuSTARmission\ source 105 matched to three \Chandra\ sources, and \NuSTARmission\ sources 70 and 57 matched to two \Chandra\ sources. In these instances, we kept all \Chandra-detected sources in our total list of 64 X-ray sources. When comparing \NuSTARmission\ classifications with optical counterpart types, we only used the optical counterpart associated with the \Chandra\ source with the largest number of counts in the 0.35-8.0 keV band.
\par
\begin{figure}
    \centering
    \includegraphics[width=0.48\textwidth,keepaspectratio]{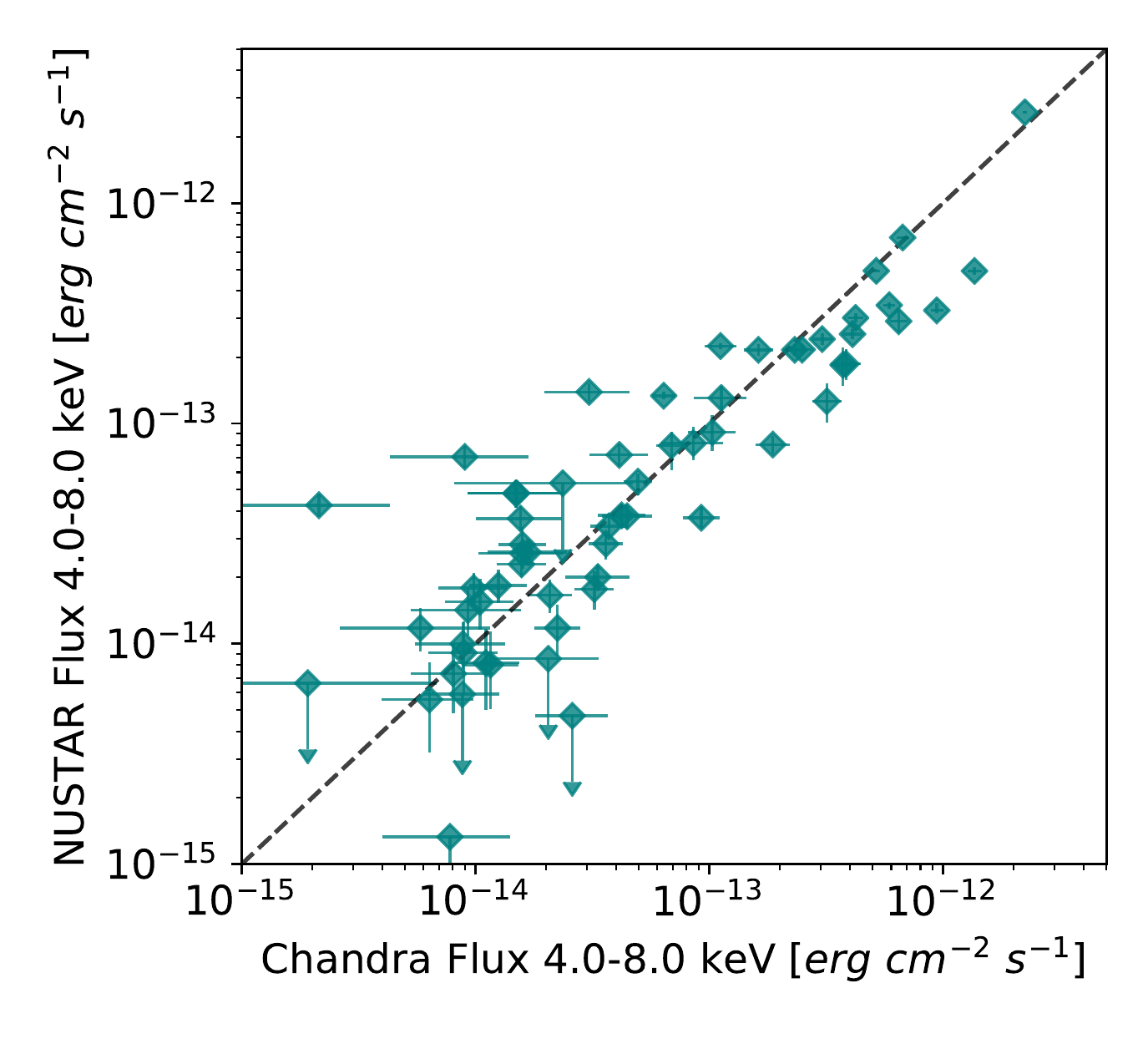}
    \caption{A comparison of \NuSTARmission\ and \Chandra\ measured fluxes for 60 hard X-ray sources observed by both telescopes. Sources were matched positionally within 10$^{\prime \prime}$}.
    \label{flux_comp}
\end{figure}

 To further confirm associations between the \NuSTARmission\ and \Chandra\ sources, we compared the flux of each match in the 4-8 keV energy range, as shown in Figure \ref{flux_comp}. We converted from count rates to fluxes for each telescope using the NASA High Energy Astrophysics Science Research Archive Center's (HEASARC) web-based Portable, Interactive Multi-Mission Simulator (WebPIMMS) tool.\footnote{https://heasarc.gsfc.nasa.gov/cgi-bin/Tools/w3pimms/w3pimms.pl}
\subsection{Identifying Optical Counterparts in PHAT}\label{opt_counterparts}
We determined optical counterparts to X-ray sources using a combination of positional matching, UV magnitude cuts, and visual inspection. We initially determined optical counterpart candidates by looking at both optical and UV images of the PHAT data within the 1-$\sigma$ \Chandra\ positional error circles of a hard X-ray source. This method allowed for initial detection of likely counterparts such as background galaxies and globular clusters \citep[e.g.,][]{Bologna}. If a source had a clear point source in the UV F336W image within the 1-$\sigma$ \Chandra\ positional errors, it was noted as a point source counterpart candidate and its PHAT photometry was retrieved and is listed in Table \ref{counterpart_table}.
\par We investigated the false match probability for the PHAT counterparts. The PHAT survey area is divided into 23 ``bricks'' (see \citet{Dalcanton} for description of brick boundaries). We calculated the source density of O/B stars in the PHAT survey in the bricks (9 bricks total) covered by \Chandra\ and \NuSTARmission\ observations. We used only stars with good data in at least three of the HST photometric bands used by PHAT and an F336W magnitude of less than 23. Dividing the number of O/B stars by the total area of the 9 bricks gives a density of O/B stars per area. We then multiply this source density by the area of the average \Chandra\ 1-$\sigma$ error circle to determine the probability of finding an O/B star within the 1-$\sigma$ error circle of an X-ray source. We find a false match probability of about 2\%. Accounting for the 64 X-ray sources in our combined \NuSTARmission-\Chandra\ sample within the PHAT footprint, we expect 1-2 false matches.

\begin{deluxetable*}{lllllllllll}
\tablecaption{PHAT photometry for HMXB optical counterpart candidates\label{counterpart_table}}
\tablehead{
\colhead{Catalog Name} &
\colhead{\Chandra} &
\colhead{\Chandra} &
\colhead{PHAT} &
\colhead{PHAT} &
\colhead{F275W} &
\colhead{F336W} &
\colhead{F475W} &
\colhead{F814W} &
\colhead{F110W} &
\colhead{F160W} \\
\colhead{} &
\colhead{RA} &
\colhead{Dec} &
\colhead{RA} &
\colhead{Dec} &
\colhead{} &
\colhead{} &
\colhead{} &
\colhead{} &
\colhead{} &
\colhead{}
}
\startdata
004240.31+411845.3 & 10.667979 & 41.312682 & 10.667543 & 41.312546 & $22.03\pm0.04$ & $21.31\pm0.01$ & $20.940\pm0.003$ & $19.799\pm0.003$ & $19.263^\pm0.003$ & $18.512\pm0.003$ \\
004249.22+411815.5 & 10.70509 & 41.304395 & 10.705035 & 41.304478 & \nodata & $25.2\pm0.2$ & $22.119\pm0.006$ & $19.314\pm0.002$ & $18.552\pm0.002$ & $17.586\pm0.002$ \\
004308.62+411248.0 & 10.785948 & 41.213448 & 10.785837 & 41.213419 & $23.19\pm0.07$ & $22.97\pm0.03$ & $23.41\pm0.01$ & $23.09\pm0.02$ & $23.26\pm0.04$ & $22.63\pm0.05$ \\
004316.10+411841.2 & 10.817115 & 41.311543 & 10.817102 & 41.31151 & $24.1\pm0.1$ & $23.15\pm0.04$ & $24.43\pm0.02$ & $22.15\pm0.01$ & $21.43\pm0.01$ & $19.906\pm0.007$ \\
004321.07+411750.2 & 10.837815 & 41.297389 & 10.837831 & 41.297466 & $24.9\pm0.2$ & $24.04\pm0.06$ & $24.43\pm0.02$ & $24.48\pm0.06$ & $25.0\pm0.2$ & \nodata \\
004321.48+411556.5 & 10.839501 & 41.265805 & 10.839394 & 41.265853 & $23.8\pm0.1$ & $23.20\pm0.05$ & $23.79\pm0.01$ & $23.19\pm0.02$ & $22.90\pm0.03$ & $21.72\pm0.02$ \\
004335.91+411433.0 & 10.899635 & 41.2426 & 10.899774 & 41.242598 & $22.62\pm0.05$ & $22.80\pm0.03$ & $24.04\pm0.01$ & $23.53\pm0.02$ & $22.97\pm0.04$ & $23.29\pm0.09$ \\
004339.06+412116.7 & 10.912737 & 41.354885 & 10.912961 & 41.354867 & $23.8\pm0.1$ & $23.23\pm0.04$ & $23.87\pm0.01$ & $23.63\pm0.03$ & $23.60\pm0.04$ & $23.27\pm0.06$ \\
004350.76+412117.4 & 10.961516 & 41.355033 & 10.961508 & 41.355045 & $22.78\pm0.06$ & $21.16\pm0.01$ & $21.694\pm0.004$ & $19.843\pm0.002$ & $19.129\pm0.002$ & $18.252\pm0.002$ \\
004404.75+412126.5 & 11.019799 & 41.35756 & 11.019758 & 41.357577 & $23.51\pm0.09$ & $22.36\pm0.02$ & $22.150\pm0.005$ & $19.938\pm0.003$ & $18.735\pm0.002$ & $17.790\pm0.002$ \\
004425.73+412241.8 & 11.107221 & 41.378442 & 11.107175 & 41.378477 & $26.1\pm0.6$ & $24.9\pm0.1$ & $24.65\pm0.02$ & $22.62\pm0.01$ & $21.880\pm0.009$ & $21.45\pm0.01$ \\
004448.13+412247.4 & 11.200545 & 41.379973 & 11.200584 & 41.380057 & $23.48\pm0.08$ & $23.39\pm0.04$ & $24.59\pm0.02$ & $22.95\pm0.01$ & $21.851\pm0.009$ & $20.448\pm0.006$ \\
004518.39+413936.0 & 11.326586 & 41.66018 & 11.326624 & 41.660177 & $25.6\pm0.4$ & $24.8\pm0.1$ & $25.28\pm0.03$ & $24.70\pm0.04$ & $24.10\pm0.05$ & $23.27\pm0.05$ \\
004527.89+413904.9 & 11.366179 & 41.651539 & 11.36618 & 41.651542 & $24.3\pm0.2$ & $22.81\pm0.03$ & $23.405\pm0.009$ & $20.870\pm0.004$ & $19.971\pm0.003$ & $18.819\pm0.002$ \\
004528.29+412943.4 & 11.367681 & 41.495538 & 11.367687 & 41.495535 & $19.76\pm0.01$ & $19.202\pm0.005$ & $20.231\pm0.002$ & $19.014\pm0.002$ & $18.526\pm0.001$ & $17.754\pm0.001$ \\
\enddata
\tablecomments{PHAT photometry for all point source optical counterparts to NuSTAR hard X-ray sources. Sources are identified by their \Chandra\ catalog name, which corresponds to the \textit{\Chandra\ Catalog Name} column in Table \ref{chan_nustar}. Ellipses indicate that the source was not detected in that filter.}
\end{deluxetable*}

\par
\begin{figure}
    \centering
    \includegraphics[width=0.5\textwidth,keepaspectratio]{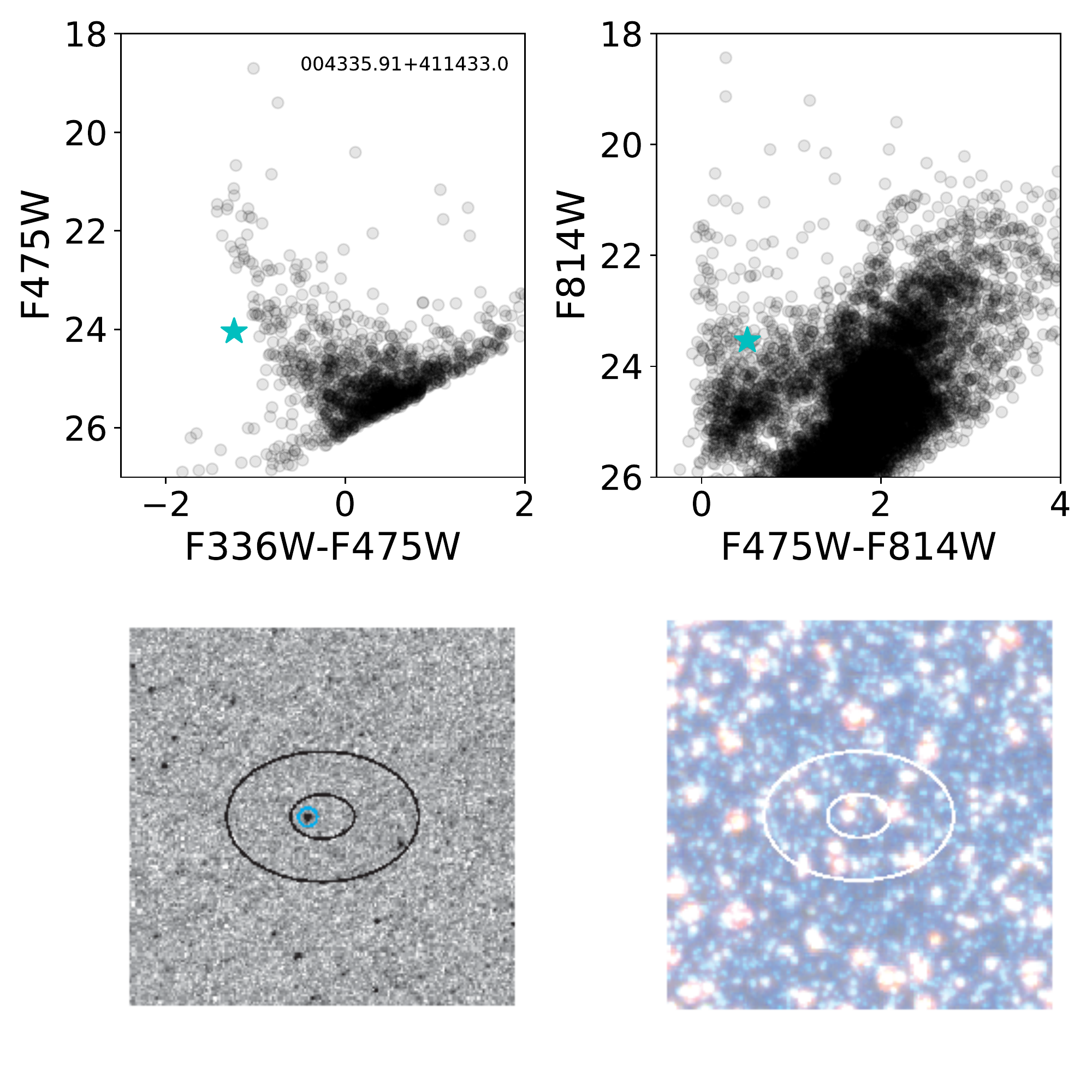}
    \caption{Color-magnitude diagrams (CMDs) (top) and finding charts (bottom) from the PHAT data set for the 10$^{\prime \prime} \times$ 10$^{\prime \prime}$ region surrounding source 004335.91+411433.4. The top two panels show the UV and optical CMDs. The bottom two panels show a UV image in the F336W filter and an RGB optical image with the F160W filter as red, F814W as green, and F336W as blue. The plotted ellipses represent the 1- and 3-$\sigma$ \Chandra\ positional errors. The optical counterpart is identified in the UV finder (lower left) with a cyan circle and on the CMDs with a cyan star. The black points in the background represent other stars in the PHAT photometry catalog within 5$^{\prime \prime}$ of the X-ray source position.}
    \label{cmd}
\end{figure}

 We used finding charts and CMDs to identify optical counterparts. Figure \ref{cmd} shows a representative figure for source 004335.91+411433.4, an X-ray source with a point source optical counterpart. The optical counterpart is marked in the UV image (lower left) with a cyan circle and is visible as a very bright star in the 1-$\sigma$ \Chandra\ positional errors of the optical finder (lower right). The counterpart is also plotted on two color-magnitude diagrams (CMDs) in the top row of the figure as a cyan star. It falls along the massive end of the main sequence in both CMDs. Note that there are far fewer stars in the upper left CMD because there are not as many stars in the PHAT survey that have well-measured UV (F336W) magnitudes. The very populated region of the upper right CMD is the red giant branch, which is too faint in the UV to be detected in the PHAT data, so that feature is not as prominent in the UV CMD.
\par We looked for optical counterparts for the \NuSTARmission-\Chandra\ sources, 64 of which are within the area of M31 observed by PHAT. We determined the following optical counterparts: 15 point sources, 13 globular clusters, and 8 background galaxies. The remaining 28 \NuSTARmission-\Chandra\ sources do not have clear optical counterparts. Optical counterparts for all sources are listed in the last column of Table \ref{chan_nustar}.
\par We expect to find 7-8 background AGN in our \NuSTARmission\ sample. Wik et al. (2018, in prep.) identified \NuSTARmission\ sources with luminosities greater than $\sim 2 \times 10^{36}\ erg\ s^{-1}$, and used the published log(N)-log(S) relationship from \cite{Harrison2016} to calculate the expected contamination of background AGN. We scale this relation to the area of the \NuSTARmission\ field also covered by \Chandra\ and \Hubble. We identify 8 background galaxies using the PHAT imaging (listed in Table \ref{chan_nustar}), which is consistent with this prediction, suggesting that all AGN with \NuSTARmission\ detections were visible in the optical PHAT data.
\section{Results}\label{sec4}
\subsection{\NuSTARmission\ Source Classification}\label{nustar_classification}
\par Wik et al. (2018, in prep.) classified the hard X-ray sources in this sample by comparing their X-ray colors and luminosities to those of Galactic XRBs with known compact object types. This method is presented in Zezas et al. (2018, in prep.) and has previously been used to classify sources in NGC 253 \citep{Wik2014}.
\par Black-hole XRBs are known to exhibit different accretion states which are manifested by their different broad-band X-ray spectra (especially above 10 keV) and power-spectra \citep[e.g.,][]{Remillard&McClintock,Done2004}. The main differences between these spectral states are identified at energies above 8.0 keV, i.e. energies that can be probed with \NuSTARmission.
\par In order to develop a diagnostic tool that can be used to characterize \NuSTARmission\ observations of extragalactic XRBs, \Zezas\ used the extensive library of black-hole spectra of \citep{Sobolweska2009}.  This library includes includes a set of 1772 Rossi-XTE - PCA observations of 6 Galactic black-hole X-ray binaries. These observations were performed during different accretion states, and in some cases they cover the complete evolution of a system during an outburst. Each spectrum was modeled with a Comptonized disk black-body model \citep{Sobolweska2009}. The state characterization was based on the spectral shape \citep[see][for more details]{Sobolweska2009}.
\par Based on this model and the \NuSTARmission\ response files, Zezas et al. (2018, in prep.) simulated \NuSTARmission\ observations and calculated the expected count-rates in different bands. Extensive tests showed that hardness ratios involving the 4.0-6.0 keV (soft), 6.0-12.0 keV (medium), 12.0-25.0 keV (hard), and 4.0-25.0 keV (full) bands give the optimal separation of spectral states, while maximizing the number of counts in each band. Luminosities and count rates of Galactic XRBs were scaled to the distance of M31 for comparison with XRBs in our sample.
\par
\begin{figure}
    \centering
    \includegraphics[width=0.5\textwidth,keepaspectratio]{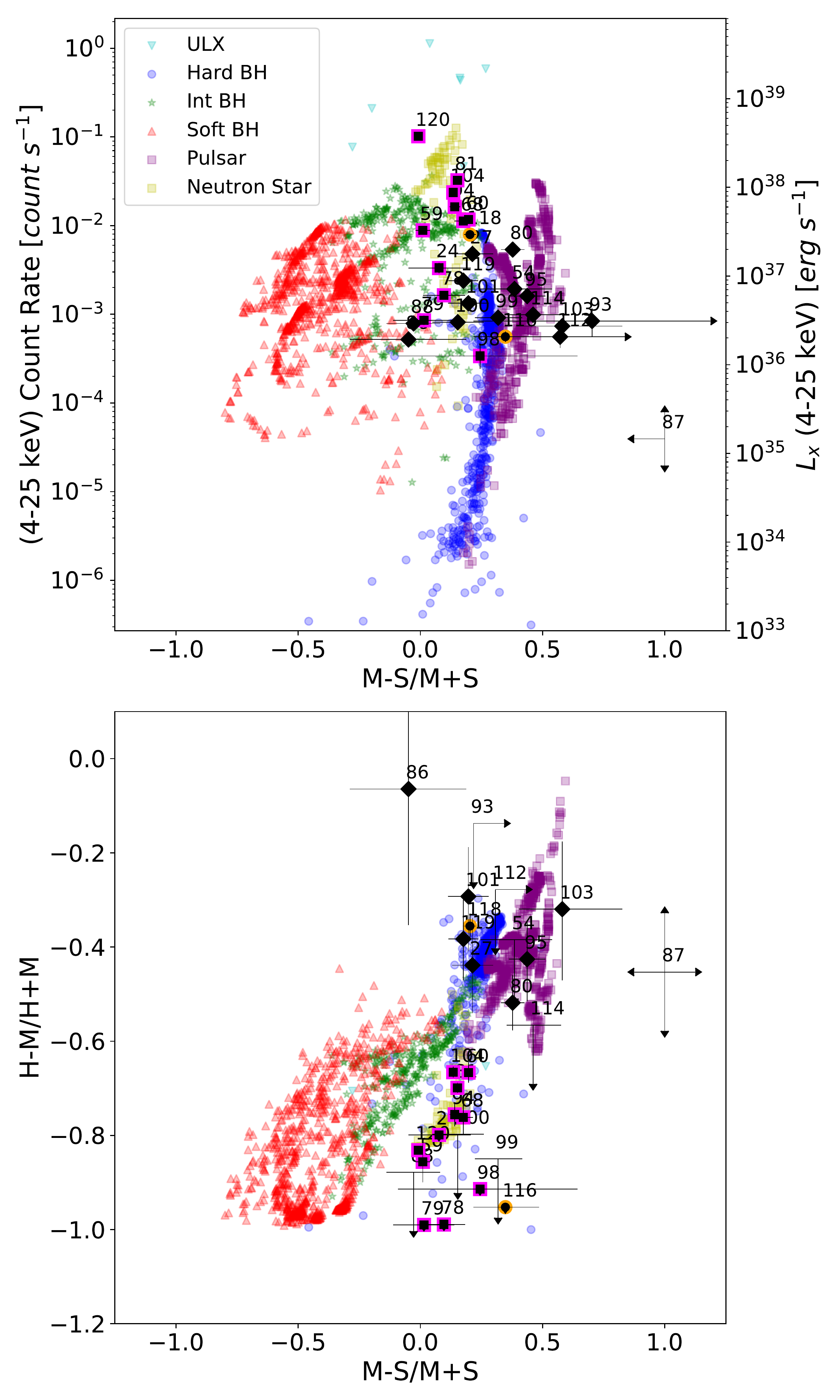}
    \caption{Hardness-intensity diagram and hardness ratio plots used to classify \NuSTARmission\ sources. These plots show sources classified as HMXB candidates as black diamonds. Circles outlined in orange are the two background galaxies with \NuSTARmission\ classifications that were identified using \Hubble\ imaging, described further in Section \ref{PHAT_Nustar}. Squares outlined in magenta are globular clusters, which occupy the non-magnetized neutron star region of both diagrams (see Section \ref{clusters} for further discussion of X-ray sources in clusters). The bands are defined as follows: soft (S=4-6 keV), medium (M=6-12 keV), and hard (H=12-25 keV). Sources are labeled by \NuSTARmission\ ID. Background colored sources represent modeled evolutionary tracks of Galactic X-ray binaries with known compact object types, adjusted for the distance of M31 (see A. Zezas et al., 2018 in prep for information on the Galactic XRB diagnostic regions). \NuSTARmission\ data and source classifications for M31 sources from Wik et al. (2018, in prep.).}
    \label{colorcolor}
\end{figure}

 We note that the highest energy of the \NuSTARmission\ data (25 keV) is well within the range of the RXTE-PCA spectra, ensuring high-quality input spectral models.  Fig. \ref{colorcolor} shows the locus of the different black-hole accretion states on the intensity-hardness ratio and hardness-ratio hardness-ratio diagrams (red, green, and blue correspond to the soft, intermediate, and hard accretion states respectively).
\par In these diagrams we also include accreting B\emph{e}-XRB pulsars with available RXTE-PCA spectra \citep[e.g.,][]{Reig2011} following the same procedure as for the black-hole X-ray binaries. Their intrinsically hard X-ray spectra clearly separate them even from the locus of the hard-state black hole X-ray binaries (Zezas et al., 2018, in prep). Finally, we include spectra of Z-track neutron star Low-mass X-ray binaries (LMXBs).
\par Determining background AGN contamination is difficult as their hard X-ray colors and luminosities can be similar to those of compact objects in the disk of M31 \citep[see e.g.,][]{Tozzi2006}. While most background galaxies do not have \NuSTARmission\ classifications because they do not have enough counts in the hard (12-25 keV) band, they can occupy similar regions of the hardness ratio and hardness-intensity diagrams as compact objects in M31. Two sources that were determined to be background galaxies using PHAT imaging are plotted as black circles outlined in orange in Figure \ref{colorcolor}. This highlights the importance of incorporating data at optical wavelengths to remove these sources from our hard X-ray sample. 

\subsection{SED Fitting of Stellar Optical Counterparts}\label{beast_section}
We obtained SED fits of 15 point source optical counterparts using the Bayesian Extinction And Stellar Tool (BEAST) \citep{BEAST}. The BEAST code fits the observed SED of an individual star in M31 with theoretical SEDs from the Padova stellar evolution models \citep{Marigo2008} using a Bayesian statistical approach. The code assumes single-star evolution and that sources are in M31. Photometric bias and uncertainty are applied from artificial star tests performed on the data. The input for the BEAST code is the six-band photometry and artificial star tests of the star measured by the PHAT survey. The code uses upper limits as constraints. Output parameters include several primary and derived quantities. Primary fit outputs include initial stellar mass, $A(V)$ (dust extinction), and stellar metallicity. Derived quantities include luminosity, effective temperature, and stellar surface gravity. Output physical parameters for the 15 point source optical counterpart candidates in our sample are listed in Table \ref{beast_table}. 
\par As part of the fitting, $\chi^{2}$ values are computed assuming multi-variate Gaussian uncertainties, either uncorrelated or correlated. The probability of a given model is proportional to $\chi^{2}$, letting us use the $\chi^{2}$ value as a relative assessment of the ``goodness of fit''.
\par In most HMXBs, the donor star is much brighter than the accretion disk at optical wavelengths, so the fits with low $\chi^{2}$ values should be robust. The BEAST code is designed to fit individual stars, and so it will return a poor fit if a point source is not an individual star. Examples of systems that might return a poor fit include parts of a multiple star system, background AGN, companions in XRBs that have been irradiated by their associated X-ray source \citep{Phillips}, stars contaminated with light from the compact object's accretion disk, chance superpositions of sources, or B\emph{e} star donors with a red excess from the accretion disk relative to the underlying B star's spectrum.

\setlength{\tabcolsep}{5pt}
\begin{deluxetable*}{llllllll}
\tablecaption{Output best fit parameters for stellar optical counterparts from BEAST SED fitting code\label{beast_table}}
\tablehead{
\colhead{Catalog Name} &
\colhead{log(L)} &
\colhead{log(g)} &
\colhead{log(T)} &
\colhead{$A_{V}$} &
\colhead{Mass} &
\colhead{$\chi^{2}$} &
\colhead{Best-fit Spectral Type} \\
\colhead{} &
\colhead{[$L_{\odot}$]} &
\colhead{[$cm\ s^{-2}$]} &
\colhead{[K]} &
\colhead{} &
\colhead{[$M_{\odot}$]} &
\colhead{} &
\colhead{}
}
\startdata
004240.31+411845.6 & $4.5^{+0.1}_{-0.1}$ & $2.2^{+0.1}_{-0.1}$ & $4.03^{+0.02}_{-0.02}$ & $2.0^{+0.1}_{-0.1}$ & $13^{+1}_{-1}$ & 49 & $\chi^{2}$ too high to trust fit\\
004249.22+411815.8 & $3.8^{+0.1}_{-0.1}$ & $0.7^{+0.2}_{-0.3}$ & $3.61^{+0.02}_{-0.03}$ & $0.7^{+0.3}_{-0.4}$ & $5^{+2}_{-3}$ & 11 & possible He burning star\\
004308.63+411248.4 & $3.1^{+0.5}_{-0.3}$ & $3.6^{+0.2}_{-0.1}$ & $4.2^{+0.1}_{-0.1}$ & $0.7^{+0.5}_{-0.3}$ & $5^{+2}_{-1}$ & 1 & B \\
004316.11+411841.5 & $5.7^{+0.3}_{-1.1}$ & $3.8^{+0.3}_{-0.8}$ & $4.6^{+0.1}_{-0.4}$ & $4.3^{+0.2}_{-0.3}$ & $45^{+34}_{-35}$ & 50 & high $\chi^{2}$, flat SED\\
004321.08+411750.6 & $3.0^{+0.5}_{-0.4}$ & $4.1^{+0.2}_{-0.3}$ & $4.2^{+0.1}_{-0.1}$ & $1.2^{+0.4}_{-0.5}$ & $5^{+2}_{-1}$ & 6 & B\\
004321.48+411556.9 & $3.6^{+0.5}_{-0.4}$ & $4.1^{+0.3}_{-0.2}$ & $4.4^{+0.1}_{-0.1}$ & $1.2^{+0.2}_{-0.2}$ & $7^{+4}_{-3}$ & 1 & B \\
004335.91+411433.4 & $4.0^{+0.3}_{-0.6}$ & $4.2^{+0.2}_{-0.3}$ & $4.5^{+0.1}_{-0.1}$ & $1.7^{+0.2}_{-0.2}$ & $10^{+3}_{-4}$ & 5 & B\\
004339.06+412117.6 & $3.4^{+0.4}_{-0.4}$ & $4.1^{+0.2}_{-0.2}$ & $4.3^{+0.1}_{-0.1}$ & $1.0^{+0.2}_{-0.2}$ & $6^{+3}_{-2}$ & 5 & B\\
004350.76+412118.1 & $6.5^{+0.1}_{-0.1}$ & $3.8^{+0.1}_{-0.1}$ & $4.72^{+0.02}_{-0.02}$ & $3.2^{+0.1}_{-0.1}$ & $106^{+8}_{-7}$ & 1756 & $\chi^{2}$ too high to trust fit\\
004404.75+412127.2 & $5.2^{+0.1}_{-0.1}$ & $1.9^{+0.1}_{-0.1}$ & $4.06^{+0.02}_{-0.02}$ & $4.4^{+0.1}_{-0.1}$ & $19^{+2}_{-2}$ & 459 & $\chi^{2}$ too high to trust fit\\
004425.73+412242.4 & $4.6^{+0.6}_{-0.7}$ & $3.4^{+0.5}_{-0.3}$ & $4.4^{+0.2}_{-0.2}$ & $3.8^{+0.2}_{-0.2}$ & $14^{+11}_{-8}$ & 8 & B\\
004448.13+412247.9 & $5.7^{+0.1}_{-0.1}$ & $3.9^{+0.1}_{-0.1}$ & $4.63^{+0.02}_{-0.02}$ & $4.6^{+0.1}_{-0.1}$ & $49^{+4}_{-3}$ & 940 & high $\chi^{2}$, flat SED\\
004518.38+413936.6 & $2.8^{+0.5}_{-0.4}$ & $4.1^{+0.2}_{-0.2}$ & $4.2^{+0.1}_{-0.1}$ & $1.4^{+0.6}_{-0.6}$ & $4^{+2}_{-1}$ & 10 & B\\
004527.88+413905.5 & $6.5^{+0.1}_{-0.1}$ & $3.8^{+0.1}_{-0.1}$ & $4.70^{+0.03}_{-0.04}$ & $5.2^{+0.1}_{-0.1}$ & $158^{+39}_{-25}$ & 667 & $\chi^{2}$ too high to trust fit\\
004528.24+412943. & $5.4^{+0.1}_{-0.1}$ & $2.5^{+0.1}_{-0.1}$ & $4.28^{+0.02}_{-0.02}$ & $2.8^{+0.1}_{-0.1}$ & $26^{+2}_{-2}$ & 3409 & $\chi^{2}$ too high to trust fit\\
\enddata
\tablecomments{Output best fit parameters for stellar optical counterparts using the BEAST SED fitting code for all point sources listed in Table \ref{counterpart_table}. Median values $\pm 33\%$ are listed. After the posterior distribution is complete, the BEAST code calculates the $\chi^{2}$ value for the most likely model, listed here. Fits are considered robust for $\chi^{2}<12$. The \textit{Probable Spectral Type} column lists the most likely spectral type of each source, based on its best fit physical properties. See Section \ref{beast_section} for a more detailed description of BEAST SED fitting.}
\end{deluxetable*}

\par
\begin{figure*}
    \centering
    \includegraphics[width=0.9\textwidth,keepaspectratio]{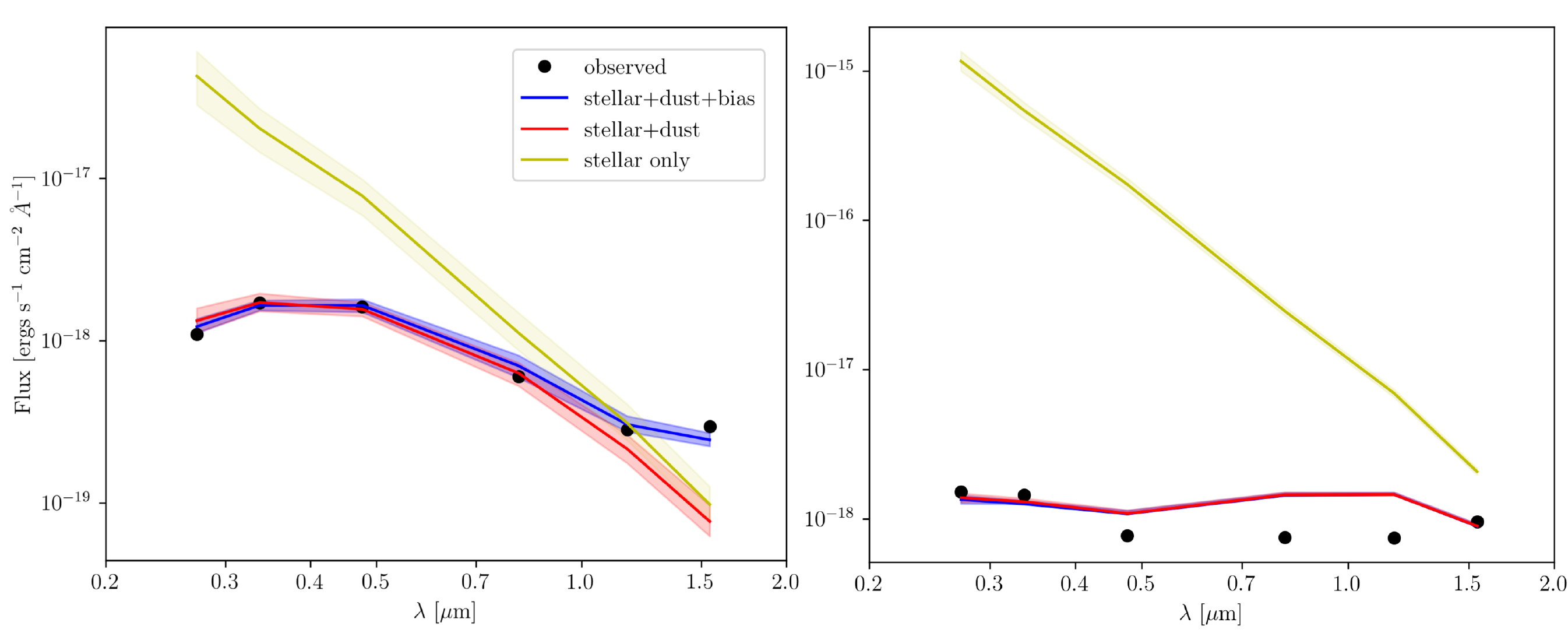}
    \caption{BEAST SED fits for the optical counterparts of two HMXB candidates, 004321.48+411556.9 (left) and 004448.13+412247.9 (right). The counterpart in the left panel has a robust fit and is most likely a B-type star. The counterpart in the right panel returns a poor fit. Black points show measured photometry for the optical counterparts from the PHAT dataset, listed in Table \ref{counterpart_table}. The colored lines show the median fit $\pm33\%$ errors of the three different models. Yellow shows the stellar-only model, red shows a stellar+dust model and blue shows a stellar+dust+bias model. The observational bias is determined using artificial star tests.}
    \label{beast_plot}
\end{figure*}

 Table \ref{beast_table} shows a clear division in $\chi^{2}$ values: $\chi^{2} \lesssim 12$ or $\chi^{2} \gtrsim 50$. Examples of these two categories are shown in Figure \ref{beast_plot}. The lower $\chi^{2}$ fit appears similar to a stellar SED model, while the high $\chi^{2}$ appears to have a flat SED. Based on the clear division in fit quality as well as SED appearance, we decided that fits with $\chi^{2} \lesssim 12$ likely have SEDs consistent with stars in M31 fits with higher $\chi^{2}$ values do not. Thus, we did not determine a spectral type for sources with $\chi^{2}$ values above $\sim 12$. 
\par Table \ref{beast_table} lists the probable spectral type given the best-fit physical parameters. We determine masses, temperatures, and luminosities for 7 of the 15 point sources which are consistent with a B-type star, and therefore very strong HMXB candidates. B-type stellar classification was determined for $4 M_{\odot} \lesssim M \lesssim 17 M_{\odot}$ and $4.0\ K \lesssim log(T_{eff}) \lesssim 4.5\ K$ \citep[e.g.,][]{Silaj2010}. Figure \ref{beast_plot} shows the SED fit for the optical counterpart to 004321.48+411556.9, an example of a good fit for a B-type star. 
\par One optical counterpart (004249.22+411815.8) is classified as a possible red Helium burning star given its high luminosity, low temperature and low surface gravity. The hard X-ray source associated with this optical counterpart does not have a \NuSTARmission\ classification.
\par We do not rule sources out as HMXB candidates due to fits because a poor fit may be returned for stars that have been irradiated by their associated X-ray source or contaminated with light from their compact object's accretion disk, as discussed previously in this section. However, sources with good fits to stars in M31 may be stronger than those that do not.
\par We find two point sources that have relatively flat SEDs, noted in Table \ref{beast_table}. Figure \ref{beast_plot} illustrates the BEAST fit for the optical counterpart of one of these sources, 004448.13+412247.9. The BEAST attempts to fit the point source as a hot star with a high A(V) (e.g., 4.6 magnitudes of extinction, as shown in Figure \ref{beast_plot}, compared to the 1.2 magnitudes of dust extinction shown for a star that has a robust fit to a B-type star) and still returns a poor fit. The other source that has an optical counterpart with a flat SED is 004316.11+411841.5. Such a flat SED may be indicative of a background AGN.
\subsection{Star Formation Histories of HMXB Candidates}\label{SFH}
\par We used the spatially resolved recent star formation history (SFH) of M31 by \citet{Lewis} to determine likely ages of HMXB candidates in our sample. \citet{Lewis} inferred these SFHs using CMDs of 100 pc by 100 pc regions in the M31 disk. 
\par We assume HMXBs contain secondary stars more massive than 7 $M_{\odot}$, which have lifetimes of 10 Myr. Thus we conservatively restrict our age distribution analysis to $<$ 60 Myr. The time resolution of the SFHs is log(time)=0.1 yr. Star formation histories are not available for regions too close to the bulge of M31 (in PHAT bricks 1 and 3) because crowding does not allow for accurate CMD fitting, and so not all HMXB candidates are included in our analysis. For that reason, 8 of the 15 HMXB candidates are used in the SFH analysis: 004335.91+411433.4, 004350.76+412118.1,\newline 004404.75+412127.2, 004404.75+412127.2,\newline 004425.73+412242.4, 004448.13+412247.9,\newline 004518.38+413936.6, 004527.88+413905.5,\newline and 004528.24+412943.9. 
\par
\begin{figure}
    \centering
    \includegraphics[width=0.5\textwidth,keepaspectratio]{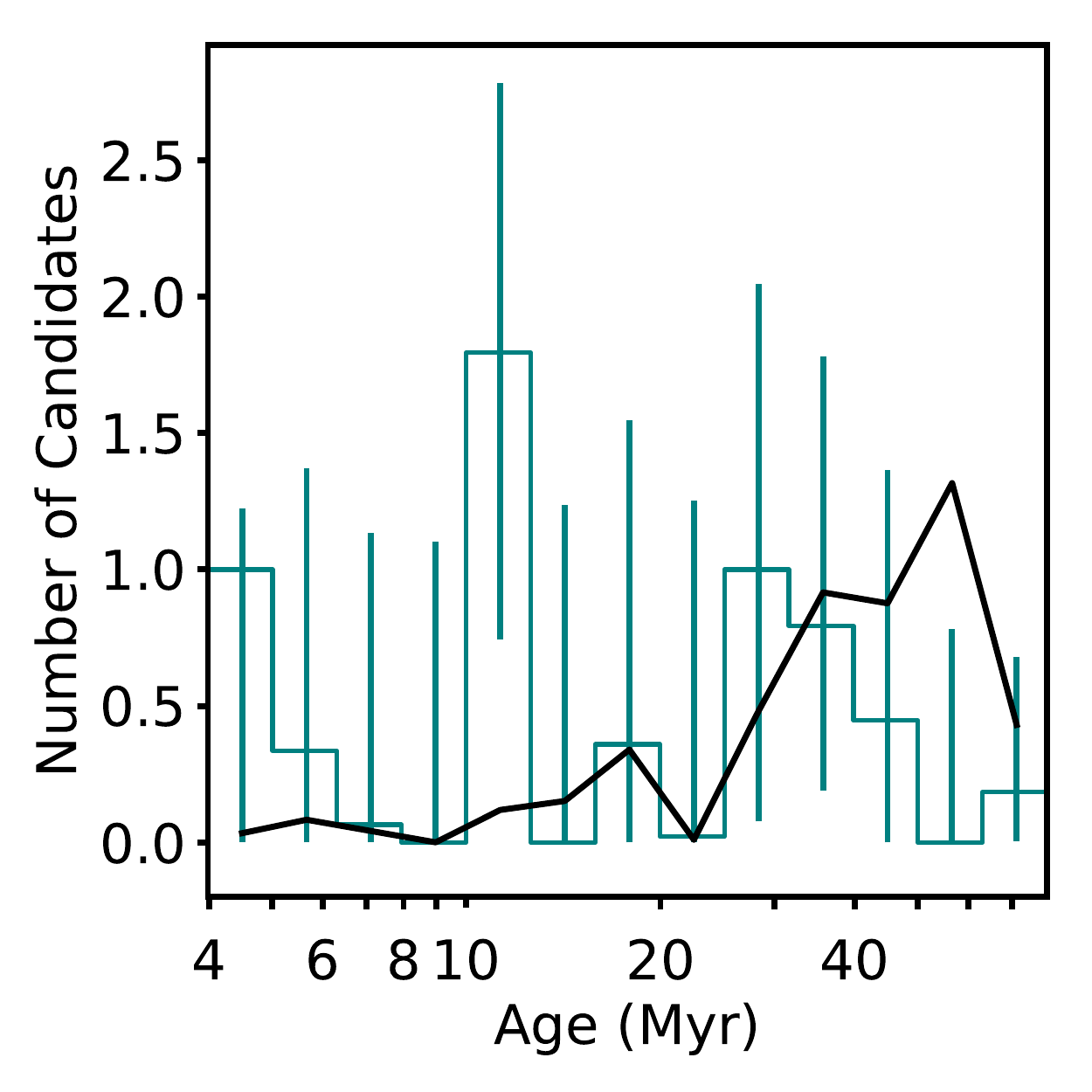}
    \caption{Histogram showing the number of HMXB candidates associated with each time bin based on their spatially resolved star formation histories from \cite{Lewis}. The time bins are defined by the age resolution of the SFHs. The black line represents the average of 100 random samples of regions in M31 not associated with HMXB candidates, providing a reference for the overall SFH of M31.}
    \label{age_plot}
\end{figure}

 For each HMXB region containing an HMXB candidate, we calculated the total stellar mass formed in the past 60 Myr. For each time bin younger than 60 Myr, we calculated the fraction of the mass formed in that bin. This fraction gives the normalized probability that the given HMXB candidate formed in that time bin. We take the uncertainties in the SFH into account by sampling the SFH 1000 times and recalculating the age distribution. We take the 16th and 84th percentile in each time bin, to determine uncertainties. The number of HMXB candidates expected to form in each time bin is shown in Figure \ref{age_plot} in teal.
\par We then compare the probability distribution for regions with HMXB candidates to the rest of the M31 disk. We do this by randomly selecting 8 regions from a sample of $\sim$ 49 regions containing known background galaxies identified by Williams et al. (2018, submitted) in the PHAT bricks observed by \Chandra\ and \NuSTARmission\ and immediately adjacent. We use the SFH for regions around background galaxies because these should be randomly distributed throughout the disk and not correlated with the HMXB population. We perform this random selection 100 times and plot the average expected number of candidates in each time bin in black in Figure \ref{age_plot}. 
\par Using the sub-sample of 8 HMXB candidates with SFHs, we were able to determine that about 3 HMXBs in our sample have an age of $\sim$25-50 Myr, 2 are $\sim$10 Myr old, and 1 is $\sim$4 Myr old. Two of the HMXB candidates analyzed, 004350.76+412118.1 and 004404.75+412127.2, are found in regions without significant star formation in the last 60 Myr, making them weaker HMXB candidates. 
\par The ages of candidates between 25 and 50 Myr are fairly consistent with random draws from the disk of M31. This indicates that the regions with these HMXB candidates do not appear to be a different age than the average population. The peak in star formation in regions surrounding HMXB candidates in the 10-12 Myr and 4 Myr time bins are more significant deviations from the overall SFH of M31, as demonstrated in Figure \ref{age_plot}. These may be probing the prompt HMXB formation channel in M31.
\section{Discussion}\label{sec5}
Our measurements allow many detailed comparisons of the X-ray sources in this region of M31.  First, we can compare the optical and X-ray characteristics of the sources with counterparts. Next, we can compare those characteristics with the age distribution of the surrounding stellar populations as an additional consistency check, and finally we can consider the sources in globular clusters to look for X-ray characteristics unique to that specific subclass. We discuss all of these comparisons below.
\subsection{Comparing \NuSTARmission\ and \Hubble\ Source Classification}\label{PHAT_Nustar}
\par
\begin{figure}
    \centering
    \includegraphics[width=0.5\textwidth,keepaspectratio]{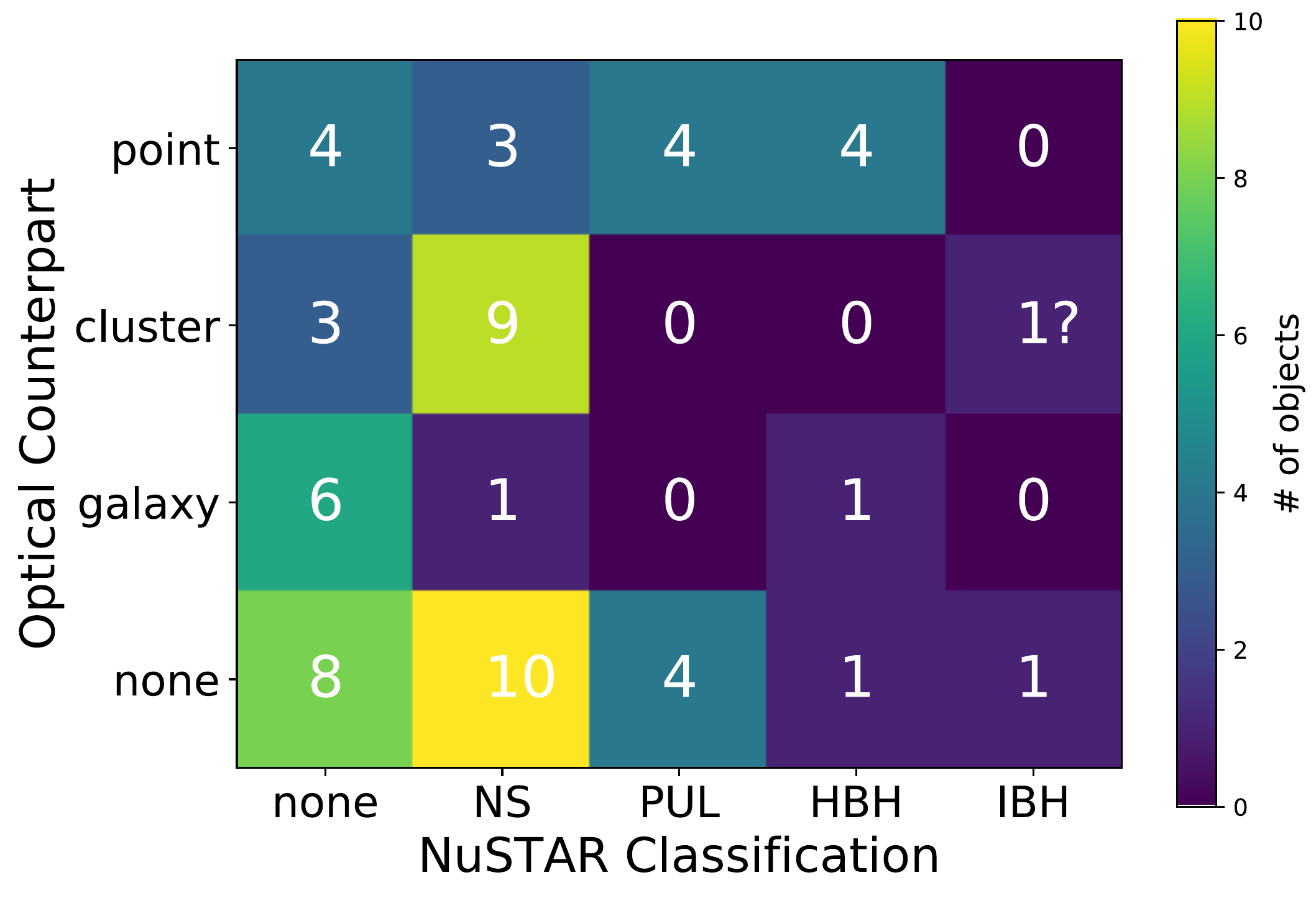}
    \caption{Comparing the \NuSTARmission\ X-ray classification and optical counterpart of sources observed with \NuSTARmission, \Chandra, and PHAT. \NuSTARmission\ classifications are defined as follows: NS=non-magnetized neutron star, PUL=pulsar, HBH=hard-state black hole, IBH=intermediate-state black hole. For further discussion of the \NuSTARmission\ classified intermediate state black hole associated with a cluster, see Section \ref{clusters}.}
    \label{class_comp}
\end{figure}

 In Figure \ref{class_comp} we compare the classification of the compact object determined by \NuSTARmission\ colors with the type of the associated optical counterpart. Sources that fall within the ``none'' \NuSTARmission\ classification did not have enough counts in all three X-ray bands to be accurately classified. Sources with the ``none'' optical counterpart classification did not have a clear optical counterpart in PHAT imaging.
\par Nine \NuSTARmission\ classified non-magnetized neutron stars in our sample are found within globular clusters, and we find no pulsars or hard state black holes in globular clusters. Roughly equal numbers of pulsars, non-magnetized neutron stars, and hard state black holes have point sources as optical counterparts. We also find that four pulsars in our sample are HMXB candidates with point source optical counterparts, while four have no optical counterpart. Pulsars without optical counterparts could be part of a low or intermediate mass X-ray binary system. This suggests that the pulsars in our sample are not preferentially in HMXB systems. When we compare the \NuSTARmission\ source classifications with the results of the BEAST SED fitting (summarized in Table \ref{HMXB}), we notice that none of the HMXB candidates with classified hard state black holes have good SED fits to B-type stars.
\par It is important to note \NuSTARmission\ sources 57 and 70, which are each blends of two \Chandra\ sources (see Section \ref{nust_chan_matching}). In both cases, one \Chandra\ source has a globular cluster optical counterpart and the other has no optical counterpart. Source 70 is classified as a non-magnetized neutron star, suggesting that the \Chandra\ source associated with a globular cluster may dominate the light detected by \NuSTARmission. Source 57 is classified as an intermediate state black hole, which would be unusual associated with a globular cluster. We compared the 4-8 keV \Chandra\ flux of the two \Chandra\ sources that matched to \NuSTARmission\ source 70 and 57. The energy fluxes were consistent within errors, so we could not determine which source dominated the flux observed by \NuSTARmission. The \NuSTARmission\ source classification may be affected by the blend of the two \Chandra\ sources, and thus additional investigation is needed to confirm if the source classification is a result of the blend.

\begin{deluxetable*}{cccc}
\tablecaption{HMXB Candidate Classification Data \label{HMXB}}
\tablehead{
\colhead{Catalog} &
\colhead{Good BEAST} & 
\colhead{SFH indicates} & 
\colhead{NuSTAR classified} \\
\colhead{Name} &
\colhead{Fit} & 
\colhead{young SF} & 
\colhead{HBH/pulsar/NS}
}
\startdata
004240.31+411845.6 &       &\nodata& \checkmark\ (hbh)\\
004249.22+411815.8 &\checkmark &\nodata& \\
004308.63+411248.4 &\checkmark &\nodata& \checkmark\ (pul)\\
004316.11+411841.5 &       &\nodata& \checkmark\ (hbh)\\
004321.08+411750.6 &\checkmark &\nodata& \checkmark\ (ns)\\
004321.48+411556.9 &\checkmark &\nodata& \\
004335.91+411433.4 &\checkmark &\checkmark& \\
004339.06+412117.6 &\checkmark &\nodata& \checkmark\ (pul)\\
004350.76+412118.1 &       &      & \checkmark\ (ns)\\
004404.75+412127.2 &       &      & \checkmark\ (hbh)\\
004425.73+412242.4 &\checkmark &\checkmark& \checkmark\ (pul)\\
004448.13+412247.9 &       &\checkmark& \checkmark\ (ns)\\
004518.38+413936.6 &\checkmark &\checkmark& \\
004527.88+413905.5 &       &\checkmark& \checkmark\ (pul) \\
004528.24+412943.9 &       &\checkmark& \checkmark\ (hbh)\\
\enddata
\tablecomments{Table evaluating likelihood of HMXB candidates based on BEAST fits, SFH, and \NuSTARmission\ classification of compact object. Check marks are given when a source is likely to be a HMXB using the given criteria: the source has a good BEAST SED fit to a stellar companion, the source is in a region with young (within the last 60 Myr) star formation, or the \NuSTARmission\ classification of the compact object is a pulsar, non-magenetized neutron star, or a pulsar. Ellipses in the SFH column indicate that no star formation history is available for the region around the source, due to crowding near the bulge of M31.}
\end{deluxetable*}

\par
\begin{figure}
    \centering
    \includegraphics[width=0.45\textwidth,keepaspectratio]{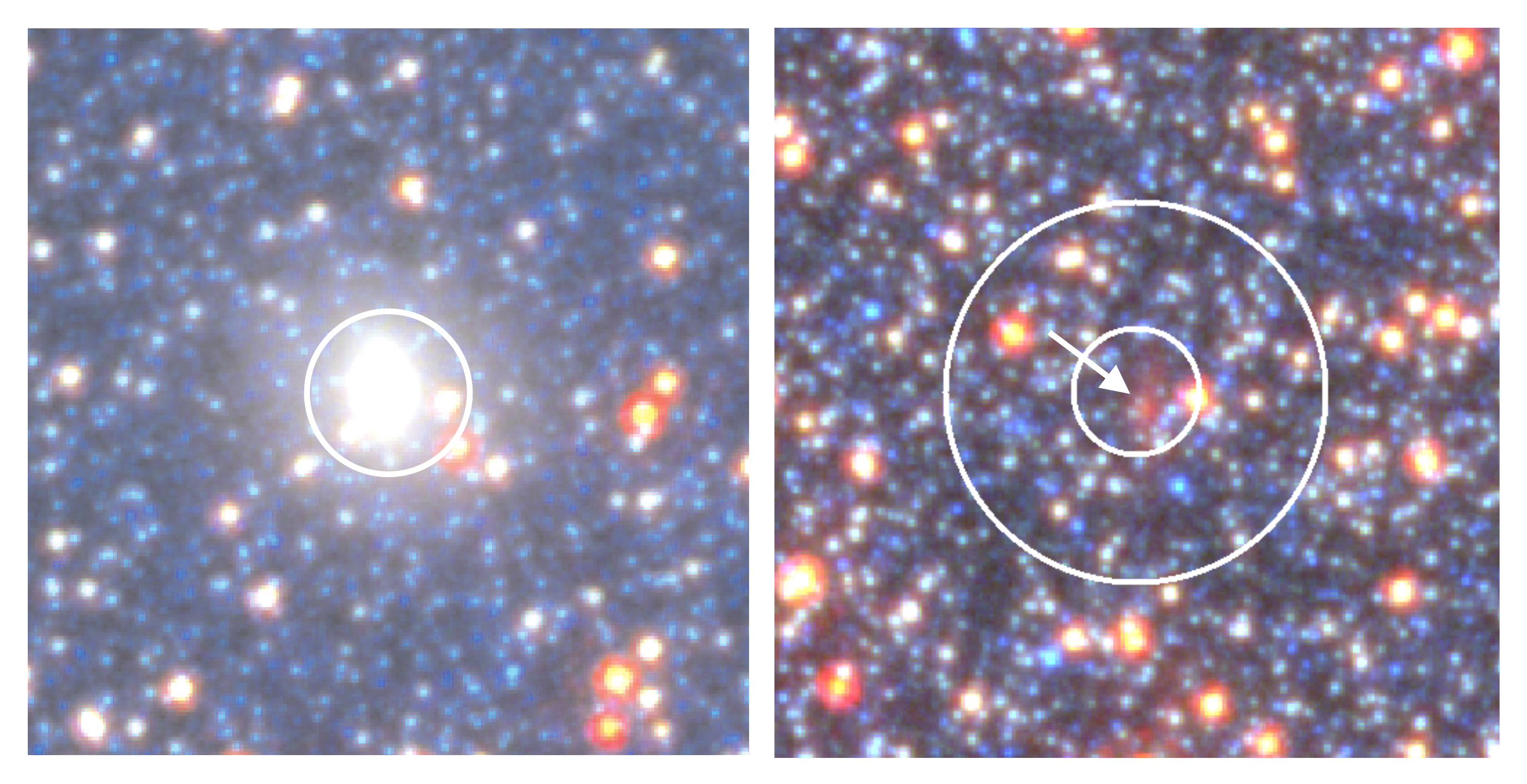}
    \caption{Images showing background galaxies associated with \NuSTARmission\ source 118 (left) and 116 (right). These color images were created with \Hubble\ imaging from the PHAT survey and use the F160W filter as red, F814W as green, and F336W as blue. Images are 10$^{\prime \prime}$ on a side. White circles indicate the 1- and 3-$\sigma$ \Chandra\ positional errors. The 1-$\sigma$ error is not visible on the image on the left as it lies on top of the bright galaxy. The galaxy on the left is bright, extended, and elliptical in shape. The galaxy on the right is much fainter. It is visible as a faint, red, extended source within the 1-$\sigma$ error error circle, indicated with an arrow. For more discussion of background galaxies, see Section \ref{PHAT_Nustar}.}
    \label{background_galaxies}
\end{figure}

 We compare PHAT imaging with \NuSTARmission\ classifications to remove background AGN contamination from our sample. We find two cases where sources classified as compact objects in the disk of M31 were determined to be background galaxies using PHAT imaging. See Figure \ref{background_galaxies} for PHAT images used to identify these background galaxies. The apparent misclassification of these two sources does not affect the conclusions of this paper, but illustrates the power and necessity of combining \NuSTARmission\ observations with \Hubble\ data to eliminate background AGN contamination. These two sources are 004527.30+413254.1 (\NuSTARmission\ source 118, discussed in more detail in Section \ref{clusters}), which is classified as a hard state black hole, and 004530.61+413600.4 (\NuSTARmission\ source 116), which is classified as a non-magnetized neutron star. Both sources have resolved background galaxies as optical counterparts in the PHAT imaging.
\subsection{Evaluation of HMXB Candidates}
\par Table \ref{HMXB} summarizes our investigation of HMXB candidates, identified by selecting for hard X-ray sources spatially coincident with UV-bright point source optical counterparts. We evaluate whether a source is a likely HMXB using three methods: (1) SED fitting with the BEAST code to determine if a massive, young star appears to be the donor, (2) age estimation using spatially resolved SFHs, and (3) compact object classification using \NuSTARmission\ hard X-ray colors and luminosities. 
\par We consider any hard X-ray source with a UV-bright point source optical counterpart an HMXB candidate, even if it does not satisfy all three criteria. For example, the optical counterpart could have a poor SED fit because of irradiation from the compact object or mass transfer, as discussed in Section \ref{beast_section}. Additionally, not having a \NuSTARmission\ compact object classification does not rule out an HMXB candidate. Sources must have enough flux in all three \NuSTARmission\ bands to be classified, so a source could remain unclassified if it is too faint or there is too much absorption to be detected in all bands. We simply comment that having a \NuSTARmission\ compact object classification or good optical companion SED fit makes an HMXB a stronger candidate because we have more information about the system.
\par We find that HMXB candidate 004425.73+412242.4 satisfies all three criteria. It has a good SED fit indicating a B-type donor star and the SFH in the region around this source shows star formation bursts within the last 60 Myr. This source's \NuSTARmission\ colors and luminosities indicate it is likely a pulsar.
\par Three HMXB candidates in our sample, \newline 004308.63+411248.4, 004321.08+411750.6, and \newline 004339.06+412117.6 have optical counterparts that have good SED fits to B-type stars and have a \NuSTARmission\ classification of pulsar or non-magnetized neutron star. However, due to their location close to the bulge of M31, spatially resolved star formation histories are not available for the region surrounding these candidates.
\par Four HMXB candidates in our sample, \newline 004448.13+412247.9, 004518.38+413936.6, \newline 004527.88+413905.5, and 004528.24+412943.9 satisfy two of the three criteria listed in Table \ref{HMXB}. All of these sources are found in regions with recent star formation but are either lacking a good SED fit to a B-type star or do not have a compact object classification from \NuSTARmission.
\par Four HMXB candidates in our sample have either a good SED fit for a B-type donor star \newline(004249.22+411815.8 and 004321.48+411556.9) or \NuSTARmission\ classified compact object \newline(004240.31+411845.6 and 004316.11+411841.5). These sources do not have SFHs available, due to their proximity to the bulge of M31. 
\par Two HMXB candidates in our sample lack good SED fits to B-type companion stars and are found in regions with no significant star formation in the last 60 Myr. These sources are indicated as HMXB candidates because they have UV-bright point sources associated with a \NuSTARmission-detected hard X-ray source. Source 004350.76+412118.1 is classified as a non-magnetized neutron star and source 004404.75+412127.2 is classified as a hard state black hole. The lack of star formation and poor SED fits could indicate that these sources are background AGN. Thus, further multi-wavelength observations are needed.
\subsection{Comparison With Other Hard X-ray Observations of M31}
\par Three sources in our sample were investigated in detail by \cite{Stiele2018} in a \NuSTARmission\ survey of the central region of M31. This study was designed to overlap with a previous \textit{XMM-Newton} survey of M31 \citep{Stiele}. \citet{Stiele2018} use hardness ratios, to classify four X-ray sources previously classified only as ``hard'' using \textit{XMM-Newton} data as X-ray binaries. Two of these new XRB candidates are in our sample: \NuSTARmission\ sources 47 and 65. We classify these sources as non-magnetized neutron stars and find no optical counterpart, suggesting that they could be LMXBs because these types of stars would be too faint to be observed in the PHAT survey. Additionally, \citet{Stiele2018} discuss several sources that are too hard to be located in the XRB area of their hardness ratio diagrams. One of these sources is in our sample (\NuSTARmission\ source 91) and is classified as a pulsar with no optical counterpart.
\subsection{Ages of Stellar Populations Hosting HMXBs}
\par The ages (20-50 Myr) of the regions surrounding most HMXB candidates in this study are consistent with the results seen in the Small Magellanic Cloud (SMC) \citep{SMC}. In the SMC, B\emph{e}/X-ray binaries are the most numerous subclass of HMXBs, and are found in regions with star bursts that occurred between 25 and 60 Myr ago. In the LMC X-ray binaries were found associated with younger regions, between 6 and 25 Myr old \citep{LMC}. The statistically significant increase in the number of HMXB candidates we found in M31 in regions with a star formation burst 10 Myr ago (Figure \ref{age_plot}) aligns with the young ages found in the LMC. 
\par We examined the 3 sources in regions with a strong peak in star formation rate in the 10 Myr bin, since this is a 1.5 $\sigma$ deviation from the background population (as shown in Figure \ref{age_plot}). The stellar population surrounding one source in particular, 004425.73+4122241.8, experienced almost all of its star formation in the 10 Myr time bin. This HMXB candidate has been classified as a pulsar with a B-type stellar companion, determined by its \NuSTARmission\ colors and BEAST SED fit. Two other sources are also located in regions with SFR peaks in this time bin: 004518.38+413936.6 and 004448.13+412247.9. Source 004518.38+413936.6 has an optical counterpart that is classified as a B-type star with no \NuSTARmission\ classification for the compact object. Source 004448.13+412247.9 is classified as a neutron star by \NuSTARmission\ but the BEAST SED fit quality is too low to determine the spectral type of the companion, but it is unlikely a single star in M31.
\par Source 004518.38+413936.6 (with a B-type stellar optical counterpart) is located in a region that also experienced significant star formation in the 4 Myr time bin. Note that connecting HMXB populations to stellar ages is important for constraining formation models of compact objects. \citet{Rappaport2005} and \citet{Schawinski2012} predict that a time delay of 10 Myr (assuming instantaneous burst of star formation) or 200 Myr (continuous star formation) may be expected between the onset of star formation and the production of X-rays, depending on star formation history. XRB pulsars have been found with similar ages in the Magellanic Clouds. \citet{Li2016} found an X-ray pulsar with an O-type counterpart star in the SMC, suggesting the system is $\sim5-6$ Myr old and \citet{Belczynski2008} find that XRB pulsars can form at ages as young as $\sim 5$ Myr. HMXBs associated with very young stellar ages (10 Myr or less) but with B-star secondaries can place a particularly important constraint on initial mass ratios of HMXBs, as such an object must have had a much more massive companion with a lifetime short enough to have become the accreting compact object.
\subsection{X-ray Sources in Clusters}\label{clusters}
\par \cite{Maccarone2016} investigated hard X-ray sources in globular clusters using combined \textit{Swift}-\NuSTARmission\ spectroscopy. Our sample of 64 X-ray sources observed by \NuSTARmission, \Chandra, and \Hubble\ includes four of the five sources in that study. These sources are not HMXB candidates as they were found to be spatially coincident with globular clusters, not point sources. We find that three of these sources (\NuSTARmission\ sources 65, 104, and 120 in our sample) are classified as neutron stars and one (\NuSTARmission\ source 118) is classified as a hard state black hole.
\par We determine the hard state black hole has a background galaxy optical counterpart rather than a globular cluster, using its PHAT imaging. We also cross-reference the globular cluster and background galaxy catalogs published by the PHAT survey and find that this source is classified as a galaxy based on its morphology \citep{Johnson2015}. 
\par This source was investigated in detail by \cite{TDW2017}, who found that it has a spectroscopic redshift, which agrees with our classification as a background galaxy. This source highlights the importance of incorporating data at optical wavelengths to remove background AGN contamination.
\par We identify one, source (NuSTAR source 57) classified as an intermediate state black hole that may be associated with a globular cluster. Some caution is warranted in interpreting this source because it is associated with two separate \Chandra\ sources, 004255.61+411834.8 and 004255.19+411835.7, and hence the NuSTAR spectrum is probably a superposition of two different source spectra. The former of the two \Chandra\ sources has a globular cluster optical counterpart, while we do not see evidence for a globular cluster associated with the latter source.
\par The potential connection between the globular cluster and a black hole is intriguing. For quite some time, it was thought that the Spitzer (1969) instability would lead to mass segregation that would, in turn, expel most or all stellar mass black holes from globular clusters from globular clusters \citep{Kulkarni1993, Sigurdsson1993}. The discoveries of strong candidate globular cluster black holes in external galaxies \citep[e.g.,][]{Maccarone2007} and in the Milky Way \citep[e.g.,][]{Strader2012,Chomiuk2013,Giesers2018} has helped motivate and support theoretical work which has shown that globular cluster may retain black holes \citep[e.g.,][]{Mackey2008,Sippel2013,Morscher2015}. 
\par The globular cluster G1 is of special interest as it has been suggested to contain an intermediate mass black hole (IMBH) on the basis of stellar dynamical evidence \citep[][for an alternative view]{Gebhardt2002,Baumgardt2003}. Its X-ray source is consistent with accretion from the putative IMBH \citep{Pooly2006}. It appeared as a detectable radio source in VLA data \citep{Ulvestad2007}, but later sensitive radio data found only deep upper limits \citep{Miller-Jones2012}, again providing an ambiguous determination of whether the cluster contains an IMBH. The IMBH classification is highly uncertain, as the X-ray observations are also consistent with emission from an LMXB \citep{Kong2010,Miller-Jones2012}. Potentially, deep \NuSTARmission\ imaging could provide some additional clues about this interesting globular cluster source as well.
\par Still, the total number of strong candidate black holes in globular clusters remains relatively small, especially at distances where the clusters' structural parameters are measurable, so \NuSTARmission\ source 57 in our sample merits follow-up work to further test the black hole hypothesis.
\section{Conclusions}\label{sec6}
In this work we present 15 HMXB candidates: hard X-ray sources observed by \NuSTARmission\ and \Chandra\ that are spatially associated with UV-bright point sources from the PHAT catalog.
\par We investigated the correlation between the \NuSTARmission\ determined compact object type and the optical counterpart determined with PHAT imaging. We find 9 \NuSTARmission\ classified non-magnetized neutron stars associated with star clusters, making this the strongest correlation in our sample, agreeing with the findings in \cite{Maccarone2016}. 
\par We did not find any pulsars or hard state black holes associated with star clusters. There did not appear to be a preference for non-magnetized neutron stars, pulsars, or hard state black holes associated with UV-bright point source optical counterparts. None of the HMXB candidates in our sample with hard state black hole compact objects have a companion star with a good SED fit to a B-type star.
\par We also find an equal number of pulsars in HMXB and LMXB systems. For the pulsars, this may point towards an interesting result, however our source statistics are too small to tell; further observations are needed. However, either the pulsars are not HMXBs, and might have intermediate donor masses such as those found in other M31 pulsar systems \citep[e.g.,][]{Esposito2016,Yukita} or perhaps their pulsar identifications are not as secure.
\par We determined likely ages for HMXB candidates using published SFHs. We find that 3 HMXBs in our sample are associated with stellar populations between 25 and 50 Myr old, and 2-3 HMXB candidates are associated with younger stellar populations: 1-2 are $\sim$10 Myr old, and 1 is $\sim$4 Myr old. These ages agree with findings in the Magellanic Clouds, M33, NGC 300, and NGC 2403. The ages we find in M31 and those found in other galaxies suggest two potential formation channels for HMXBs.
\par Beyond our results investigating individual X-ray sources, this study demonstrates the ability to study both the compact object and companion star in an XRB from the hard X-rays to the near infrared using \NuSTARmission, \Chandra, and \Hubble. In this work we were able to utilize classifications by \tNuSTAR\ of hard X-ray sources as neutron stars or black holes based on their X-ray colors and luminosities. Matching the \NuSTARmission\ sources to \Chandra\ allowed us to determine the positions of these X-ray sources with increased accuracy, and thus find and classify their optical counterparts using the PHAT data set. This study is an exciting foray into the combination of hard X-ray and deep optical observations in nearby galaxies. Given the maturity of the PHAT data set, we are able to harness the data products created by the many scientists on the PHAT team to determine ages and spectral types. 
\par We look forward to continuing this work in other local galaxies as more deep \Hubble\ and \NuSTARmission\ observations are made. We also plan to compare these observational XRB population data to the predictions of theoretical population synthesis codes \citep[e.g.,][]{Sorensen2017} to place constraints on models of the formation and evolution of these systems.
\acknowledgements We thank Antara Basu-Zych for useful discussions that led to improvement of the paper. We acknowledge funding through \Chandra\ program award  GO5-17077Z (P.I. Hornschemeier). Support for this work was provided in part by \Chandra\ Award Number GO5-16085X issued by the \Chandra\ X-ray Observatory Center, which is operated by the Smithsonian Astrophysical Observatory for and on behalf of the National Aeronautics and Space Administration under contract NAS8-03060.

\bibliographystyle{likeapj}


\end{document}